\documentclass[a4paper,onecolumn]{article}
\RequirePackage[T1]{fontenc}
\usepackage[margin=0.73in]{geometry}
\RequirePackage{graphicx}
\RequirePackage{mathptmx}      
\usepackage{amstext}
\usepackage{amssymb,mathtools}
\RequirePackage{flushend}
\RequirePackage[authoryear,comma,sort&compress]{natbib}
\RequirePackage[colorlinks,citecolor=blue,urlcolor=blue,linkcolor=blue]{hyperref}
\usepackage{microtype}
\usepackage{tikz}
\usepackage{lineno}
\usepackage{float}
\usetikzlibrary{decorations.markings}
\usetikzlibrary{arrows.meta}

\begin{document}

\title{Galactic Scaling Rules in a Modified Dynamical Model}

\author{Hossein Shenavar}

\author{Hossein Shenavar\thanks{E-mail: h.shenavar@mail.um.ac.ir}
\\
Department of Physics, Ferdowsi University of Mashhad, P.O. Box 1436, Mashhad, Iran.
}

\maketitle

\begin{abstract}
Schulz galactic scaling rules  \cite{2017ApJ...836..151S}, which include baryonic Tully-Fisher relation, have been surveyed in this work within the context of a modified dynamical model. These scaling relations   are derived by employing the virial theorem and applying equilibrium and stability conditions. The scaling rules are also obtained by dimensional analysis of an integral relation between  surface density and circular velocity of disk galaxies. To check the validity of the scaling relations based on observational data, we have defined, based on the properties of the model, the proper equilibrium size $R_{eq}$  and  equilibrium velocity $V_{eq}$ of systems. By employing these measures of length and velocity,   SPARC  data \citep{2016AJ....152..157L} is used to analyze the results. The viability of the scaling relations is tested and it is shown that, compared to some   other measures of length and velocity, $R_{eq}$ and $V_{eq}$ provide the closest fits to the theoretical predictions.  We have compared our results with prior works and have concluded that  the set of  baryonic Tully-Fisher relation plus mass-size relation ( or mass-velocity ) provides an appropriate description of the general characteristics of the systems. Lastly, it is shown that these scaling relations predict a certain evolution of galactic properties  with redshift. This  behavior provides a  chance to examine the cosmic evolution of the present modified dynamical model in future works.
\end{abstract}

\section{Introduction}
In considering each particular galaxy, one could observe that different systems display distinct features such as bars, boxes,  warps, etc. This approach of close inspection of galactic systems would  lead to an intricate  classification of galaxies. Using this method, the dynamical behavior of various systems might be understood in detail.  However,  another productive approach could be the investigation of global properties of galaxies such as luminosity, mass, velocity, size, etc. This second line of research leads to galactic scaling laws. It is reasonable to argue that accurate galactic scaling laws could be used as valuable guidelines toward better understanding the origin and evolution of galaxies. As it is explained by  \cite{2016ARA&A..54..597C}, scaling  relations provide astronomers with some easily measurable, statistically motivated, characters  of galactic systems  which are redshift dependent; thus, the results can be compared with the outcome of numerical simulations. Moreover, the evolution of the scaling parameters is indeed  expected to depend on various mechanisms of  galaxy formation; therefore, the general trend of the evolution  could be employed in studying the physics of galaxy formation.  However, a dominant challenge in deriving the scaling rules is the issue of the sources of error as it has been discussed by \cite{2011ApJ...726...77S} extensively.

One of  the most accurate scaling rules is reported by \cite{1977A&A....54..661T} as a relation between  luminosity and circular velocity. However, some researchers have provided evidence in support of a baryonic Tully-Fisher relation (BTFR) which correlates the baryonic mass with a characteristic velocity, i.e. the flat velocity, of  galactic systems \citep{1999ASPC..170....3F,2000ApJ...533L..99M,2005ApJ...632..859M,2016ApJ...816L..14L}.    The argument in favour of BTFR is that the scatter in this relation  is notably smaller in near-infrared (NIR) compared to ultraviolet or far-infrared. In addition, since NIR luminosity is usually considered as the best measure of the stellar mass, one could conclude that the Tully-Fisher relation primarily describes a link between mass and rotational velocity ($M-V$) \citep{2017ApJ...836..151S}. However, it seems that we still have a lot to learn from this scaling relation. For example,  \cite{2017MNRAS.466.4159I} show that BTFR derived from dwarf irregular galaxies of LITTLE THINGS is in excellent agreement with those of larger scale galaxies; though,  \cite{2019ApJ...883L..33M,2020MNRAS.495.3636M} report a few isolated gas-rich ultra-diffuse galaxies which deviate clearly from BTFR. See section 6 of  \cite{2017MNRAS.466.4159I} and Figure 3 of  \cite{2019ApJ...883L..33M}.

 Another scaling  rule  is proposed by   \cite{1960ApJ...131..293B} as a correlation between mass,  velocity and size ( $M-V-R$ ) of galaxies. Also,   \cite{2017ApJ...836..151S} provides four interconnected scaling relations between baryonic mass, radius and velocity of galaxies. These scaling rules are derived based on two simple assumptions - related to the phenomenology of galaxies - and they include  BTFR and  \cite{1960ApJ...131..293B} relation. By surveying the literature, one may easily find other works mentioning similar results in various systems \citep{1981MNRAS.194..809L,1995AIPC..336..495K}.

The origin of galactic scaling rules is considered to be rooted within the governing dynamics of the systems. For example, it is expected in  the context of the standard model of cosmology that systematic variation of galactic  features  with halo properties, i.e.  mass, size, etc., could generate galactic scaling relations  \citep{1997gsr..proc....3W,2018arXiv181102025N}. In this case, an essential issue is defining a suitable scenario by which one could  connect  the properties  of the dark and baryonic matter to the  pure observational characters.  \cite{2018arXiv181102025N} argues, based on N-body and hydrodynamical simulations,  that the self-similar nature of cold dark matter halo  is also of critical importance in deriving the scaling rules. In these simulations, the correlations between galaxy's mass and size with those of halo are found to be highly non-linear. See also   \cite{2019ApJ...877...64D} who survey a sample of 48 spiral galaxies and report correlations between maximum disk rotational velocity (and  halo's dark  matter ) with spiral arm pitch angle, central velocity dispersion and mass of the central black hole. See  \cite{2019A&ARv..27....2S} for a review on the distribution of dark matter in galaxies and its properties.

On the other hand, within theories of modified dynamics and modified  gravity one should be able to find the scaling relations based on the modified governing equations. These scaling rules eventually correlate only the observable galactic properties, i.e. no halo is presumed in this viewpoint. See   \cite{2010dmp..book.....S} for a valuable review on different aspects of the mass discrepancy problem, including the scaling relations.

It is quite interesting that the scaling rules derived by \cite{2017ApJ...836..151S} show a close connection to MOND (MOdified Newtonian Dynamics) phenomenology because, in deriving these relations, Schulz presumes that the  acceleration at the edge of the galactic stellar disk is proportional to the characteristic acceleration of MOND. Proposed by  \cite{Milgrom:1983ca,Milgrom:1983pn,Milgrom:1983zz}, MOND considers a modification in Newton's second law to solve the missing mass problem at galactic scales. The second law of dynamics in this theory is modified to a general form of $\vec{a}\mu(a/a_{0MOND})=\vec{g}$, in which $a$ is the acceleration, $g$ is the gravitational field, $\mu$ is an  interpolating function and  $a_{0MOND}=1.2\times 10^{-10} m/s^{2}$ is a phenomenological parameter of the model which is close to $cH_{0}$ ( Here, $c$ is the speed of light and  $ H_{0}$ is the Hubble constant ). See also \cite{2012LRR....15...10F} for an excellent review on MOND theory and phenomenology. MOND has had some very interesting predictions such as the externeal field effect \citep{2020ApJ...904...51C,2021ApJ...910...81C,2021arXiv210904745C}. Moreover, within the context of MOND, one could infer BTFR from the equation of motion \citep{Milgrom:1983pn} as well as the  scale invariance of the theory \citep{2009ApJ...698.1630M}. It is important to note that the former method relates the asymptotic speed of the system to its total mass while the latter method connects some measure of the mean squared velocity to the total mass of the object \citep{2009ApJ...698.1630M}. The interpretation of MOND phenomenology is still debated; however, this model might be discussed through a Machian viewpoint   as a type of vacuum effect \citep{1999PhLA..253..273M}.

In this work, we obtain four scaling rules within the framework of a modified dynamical model ( MOD ) which is derived by imposing Neumann boundary condition on general relativistic field equations \citep{Shenavar:2016bnk,2018arXiv181005001S}. The model, which is based on Wheeler's interpretation of Mach's principle \citep{wheeler1964mach},  is reviewed in Sec. \ref{Properties}. The scaling rules are established based on two different approaches: In the first approach, reported in Sec. \ref{GSRV},  the  scaling laws are found as a result of   "collective behavior" of the systems by forcing the  equilibrium and stability conditions based  on virial theorem. The second approach uses the modified Poisson equation to find an integral relation between the baryonic surface density and the circular velocity. Then, the scaling rules are derived based on the dimensional analysis of this integral. See \ref{Derive}. These two derivations show that the scaling relations could be obtained from the first principles  as well as averaged properties of the of the systems.

To analyze the scaling relations, we have defined  measures of mass, velocity and scale length based on characteristics of baryons in each galactic systems. Then, a  sample of galactic data known as SPARC   \citep{2016ApJ...816L..14L,2016AJ....152..157L}  is used to check the reliability of the scaling laws. Some implications of these scaling rules are discussed in Discussions \ref{Discussion} and Conclusion \ref{Conclusion}.

From data analysis point of view, if there are $n$ parameters $\left\lbrace  x_{1}, ..., x_{n}  \right\rbrace $ describing a system, then one could check for  $2^{n}-n-1$ ( $=$ number of subsets containing more than one element )  relations  between them; however, not all of these relations are independent because they involve the same quantities. For example, if we presume that there are  only three characteristics of a galactic system - namely mass $M$, velocity $V$ and radius $R$ - then there could be four correlations. These relations could be displayed as $M-V$, $M-R$, $V-R$ and finally $M-V-R$; though,    only two of the relations could be considered as independent. Because performing a plane fit when there is a collinearity problem is typically more challenging than the simple line fit  ( this issue happens in the case of $M-V-R$ relation, see Sec. \ref{Datan} ), it is wise to choose the relations containing two parameters. However, the above argument is based on the presumption that the measures of mass, scale, velocity and also the underlying model ( concerning the nature of dark matter / gravity ) are well-understood and verified. This is not the case in galactic systems; thus, we have to be careful in interpreting the results. As we will see in Sec. \ref{Measures}, there are various proposals for characteristic velocity and length of the system. The nature of mass discrepancy at galactic scale too is still debated. Thus, it is more cautious to fit all three bivariate relations. The results will be provided in Sec. \ref{Datan}.  Clearly,  sharp deviations from  theoretical predictions in any of the relations  could signal anomalies due to unsuitable measures, false theories or even new parameters ( such as the presence of dark halos ). 

\section{Properties of the model}
\label{Properties}

\subsection{Review on the Modified Dynamical Model}
The dominant current view toward gravitation is based on the assumption that the laws of local physics, presented through some differential equations, are independent of the universe at large scales ( which is mainly discussed as boundary conditions ). See  \cite{1973lsss.book.....H}, page 1. In other word, it is usually presumed that the interaction of two point masses here does not depend on the existence of matter far away. This work is based on a model which presumes otherwise. In other words, in the present model,  a connection between local and global physics is prescribed.

 The gravitational interaction is described through Einstein field equations which consist of ten partial differential equations (PDE). In this system of coupled PDEs, four equations are constraint equations ( due to Bianchi identity $\nabla^{\mu} G_{\mu \nu} = 0$ ) while the rest govern the fabric of spacetime. See  \cite{1973grav.book.....M}, page 409, for a detailed discussion. Similar to other PDEs, Einstein field equations - even in their linearized form - do not uniquely determine the geometry; i.e. one needs to provide suitable boundary conditions to derive a unique solution. For example, according to  \cite{1973PhRvD...7.3563T}, the usual demand in solving the field equations  that the scalar potentials should go to zero at spatial infinity is a type of boundary condition.

  The implementation of  boundary conditions on GR field equations had been interpreted by  \cite{wheeler1964mach} as  Mach's principle. The model that we apply here has been derived based on imposing Neumann boundary condition ( BC ) on Einstein field equations \citep{Shenavar:2016bnk,2018arXiv181005001S}. Assume the following  perturbed flat Friedmann-Lemaitre-Robertson-Walker   metric 
\begin{eqnarray}  \label{metric}
ds^{2} = -(1+2\Phi)dt^{2} + R^{2}(t)(1-2\Psi)\delta_{i j}dx^{i}dx^{j}
\end{eqnarray} 
in which the parameter  $R(t)$ is the scale factor while the scalar potentials  $\Phi $ and $ \Psi $ are, in general, functions of spacetime coordinates. Using this metric, one could derive the Einstein field equations  order by order. For example, it is easy to see that  if the anisotropic stress is negligible, then the tidal forces would be equal $\partial_{i} \partial_{j} \Phi = \partial_{i} \partial_{j} \Psi$. This equation relates the two scalar potentials and when the boundary condition is imposed on a sphere with the radius of particle horizon, then it is possible to show that due to the homogeneity and isotropy of the cosmos at large scale, the most general solution would be found as $\Phi -\Psi = c_{1}(t)$. This relation shows that, due to the presence of cosmic matter at large scale, the gravitational potential $\Phi$ and the 3-curvature perturbation $\Psi$ could in general be different.  See   \cite{2018arXiv181005001S} for detailed derivation.  

According to the present model, a nonzero Neumann parameter $ c_{1}(t)$   indicates the effect of   distant stars on local dynamics. In the language of  \cite{2013SHPMP..44..242E}, this is a type of top-down causation. The boundary, i.e. the sphere defined by the particle horizon, is itself evolving; therefore, the Neumann parameter  $ c_{1}(t)$ evolves with time and this evolution could be determined from the system of equations presented in Appendix A of   \cite{2018arXiv181005001S}. However, according to these equations, the parameter $ c_{1}(t)$ changes very slowly in recent cosmic times. Thus, when one is considering the physics of the local universe,  as we will do here, one could simply assume this parameter as a constant $ c_{1}(t) =c_{1}$.  The value of  $c_1$  should be estimated from observations \citep{Shenavar:2016bnk,Shenavar:2016xcp,shenavar2018local}. It could be shown that for a system of particles, with the density $\rho(\vec{x}^{~\prime})$ and total mass $M$, one could derive the next modified potential  \citep{Shenavar:2016xcp}
\begin{eqnarray}   \label{Potential} 
\Phi (\vec{x}) =-G \int \frac{\rho(\vec{x}^{~\prime})d^{3}\vec{x}^{~\prime}}{| \vec{x}^{~\prime} - \vec{x}|}  + \frac{2c_{1}a_{0}}{M}\int \rho(\vec{x}^{~\prime})d^{3}\vec{x}^{~\prime}|\vec{x}^{~\prime} - \vec{x}|.
\end{eqnarray}
in which  $G$ is the gravitational constant and $a_{0} \equiv cH_{0}$ is a fundamental acceleration of the model related to the expansion of the Universe. Thus,  the first term on the right hand side ( RHS ) is the Newtonian potential while the second one is due to the new boundary condition. Using this potential, it is possible to obtain the motion of objects in solar system and galactic scales. Based on observations at these scales, one could see that the values $a_{0} \equiv cH_{0}=6.59 \times 10^{-10}~ m/s^{2}$  and $c_{1}=0.065$ are compatible with the data   \citep{Shenavar:2016xcp,shenavar2018local}.    At the scale of the solar system, for example, the second term on the RHS of Eq. \ref{Potential}  suggests a small, though detectable, correction to the precession of perihelion of the objects \citep{Shenavar:2016xcp}.
 It  should be mentioned that such effect occurs as a result of coupling between the zero and the first order terms in an expanding universe. In other words, such term  could not appear  in a static universe because adding a constant of integration to the scalar potential would not change the force in the static universe.

Theories with a new linear potential, i.e. a constant acceleration or a Rindler term, have been proposed before to address the mass discrepancy problem.  For example, we could mention  fourth order conformal gravity \citep{1989ApJ...342..635M,1994GReGr..26..337M,2006PrPNP..56..340M}, five-dimensional brane world unification of space, time and velocity \citep{2003NuPhS.124..258C,1999astro.ph..7244B} and Grumiller's effective theory of gravity \citep{2010PhRvL.105u1303G,2011PhRvL.106c9901G}. Also, many works in the literature have followed this line of thought to fit the basic parameters of the modified models \citep{2012MNRAS.421.1273O,2012PhRvD..85l4020M,2013MNRAS.430..450L}.

Regarding Eq. \ref{Potential}, we point out that having a new potential term, which is proportional to the distance of the two particles $i$ and $j$, i.e.  $\propto |\vec{x}_{i}-\vec{x}_{j}|$, the long range interaction in the present model is significantly enhanced compared to the pure Newtonian one ( in which the potential  behaves as $\propto 1/|\vec{x}_{i}-\vec{x}_{j}|$ ). This would provide a critical clue in following analysis when we introduce a scale length for gravitating systems.

In the history of potential theory, a  potential similar to our modified dynamical  term, i.e. $\int \rho(\vec{x}^{~\prime})d^{3}\vec{x}^{~\prime}|\vec{x}^{~\prime} - \vec{x}|$, has been named "superpotential" by  \cite{1962ApJ...135..238C,1962ApJ...136.1032C,1962ApJ...136.1037C}. These authors surveyed the properties of this term to investigate the stability of astrophysical objects, such as Jacobi ellipsoids, by employing higher ranks of tensor virial  theorem. See also  \cite{1962ApJ...136.1108R}.

From Eq. \ref{Potential} it is convenient to derive the modified Poisson equation as
\begin{eqnarray}     \label{secondorder}
\nabla^{2} \Phi = 4 \pi G  \rho + \frac{4c_{1}  a_{0}}{M}  \int \frac{\rho(\vec{x}^{~\prime})d^{3}\vec{x}^{~\prime}}{| \vec{x}^{~\prime} - \vec{x}|}
\end{eqnarray}
while, since $\nabla^{2} \int \rho(\vec{x}^{~\prime})d^{3}\vec{x}^{~\prime}|\vec{x}^{~\prime} - \vec{x}|=-2\int \frac{\rho(\vec{x}^{~\prime})d^{3}\vec{x}^{~\prime}}{| \vec{x}^{~\prime} - \vec{x}|}$, one could obtain the next fourth-order Poisson equation \citep{shenavar2018local}
\begin{eqnarray}      \label{fourthorder}
\nabla^{4} \Phi = 4 \pi G \nabla^{2} \rho - \frac{16c_{1} \pi a_{0}}{M} \rho .
\end{eqnarray} 
In most occasions,  this fourth-order differential equation provides a more convenient starting point for the simple reason that it is a pure differential equation while Eq. \ref{secondorder} is an integro-differential equation. However, as we will see in the following,  Eq. \ref{secondorder} could also provide a good insight into the physical behavior of the model specially when we are trying to compare the results of the present model with those of the dark matter models. See Sec. \ref{rhoc1} below.

The left hand side ( LHS ) of Eq. \ref{fourthorder} appears mostly in the theory of linear elasticity and it is   known as the biharmonic equation. Any solution of Laplace equation is also a solution to the biharmonic equation; though, the opposite is not correct. See Chapter 8 of   \cite{selvadurai2013partial}    for a review on this equation, its properties and solutions.  Also, for an investigation on the existence and uniqueness of the solutions of the biharmonic equation see  \cite{bhattacharyya1988existence}. 

The potential \ref{Potential} predicts that  at outer parts of galaxies which the Newtonian term dies-off, the modified term provides a constant acceleration as $2c_{1}a_{0}$ ( neglecting the long range interaction with the environment ). This prediction has been tested by using  a sample  of 101 high surface brightness ( HSB ) and low surface brightness ( LSB ) galaxies \citep{Shenavar:2016bnk}. The value of $c_{1}=0.065$ was found to be in accordance with those derived based on rotation curve data (  a sample of 39 LSB galaxies tested in    \cite{Shenavar:2016xcp} ) and perihelion precession. See also Figure A2 in   \cite{2018arXiv181005001S} which is based on 551 objects and shows similar results.

The general behavior of the  rotation curve of a galaxy with central surface density $ \Sigma_{0} $  in the present model is dependent to the ratio of  $\mathcal{R}_{F} \equiv   \Sigma_{\dagger} / \Sigma_{0} $ which one could call the Freeman ratio \citep{shenavar2018local}. Here, $ \Sigma_{\dagger} = a_{0}/G = 9.9~ kg m^{-2} $ is a critical surface density. The value of $\Sigma_{\dagger} $,  when interpreted as a surface brightness, is close to the one found by  \cite{1970ApJ...160..811F} who  first suggested that there is a universal surface brightness for  spiral galaxies \citep{2012LRR....15...10F}. As discussed in  \cite{Shenavar:2016xcp} in more details, galaxies with low mass densities, i.e. when  $\Sigma_{0} $ is small compared to $\Sigma_{\dagger}$, show a clear rising rotation curve, those galaxies with intermediate to high-mass surface densities display flat rotation curves while galaxies with the highest masses show a Keplerian decline in their rotation curves. From observational point of view, this behaviour has been reported, for example, by  \cite{1991AJ....101.1231C}.

The local stability of stellar and fluid disks in this modified dynamical model has also been  investigated and it has been shown that systems governed by Eq. \ref{fourthorder} are  locally more  stable compared to the Newtonian model \citep{shenavar2018local}. Moreover,  it could be seen, using WKB approximation,  that  LSB  galaxies are more stable than  HSBs; thus, the central surface density of galaxies is critical  in understanding  the local stability. The role of the central surface density will be discussed more in following.

At cosmic scales, we should point out that choosing Neumann BC to solve cosmic perturbation equations would also change the trajectory of massless particles which results in a new lensing equation \citep{Shenavar:2016bnk,2018arXiv181005001S}. The lensing equation of MOD has been tested against a sample of ten strong lensing systems \citep{Shenavar:2016bnk}. While the derived mass of  nine systems are found to be within the observational bound, one lens, i.e. Q0142-100,  shows a 7.5 $\%$ deviation from the lower mass bound which is acceptable regarding uncertainties in the position of the image and specially the oblateness of the lens \citep{Shenavar:2016bnk}.  The growth of  the cosmic structures too, was studied by a Newtonian  approach for which the model predicted  a more rapid rate of the structure formation in matter dominated era. Moreover, it is proved that this modified dynamical model displays a late time accelerated expansion with  an equation of state which  converges to $w<-1$ \citep{2018arXiv181005001S}.   \cite{2019MNRAS.488.3876B} provide a comparison between  the present model and $\Lambda CDM$ ( $\Lambda ~+ $ cold dark matter ) based on angular size - redshift data.

Although  Eq. \ref{Potential} has been derived by modifying the first principles, i.e. the boundary condition of GR plus including the measurement process of spacetime intervals \citep{Shenavar:2016bnk,2018arXiv181005001S}, it represents eventually a modified dynamical model. In fact, the  behavior of the present model at galactic scales shows some similarities with MOND because both models are constructed based on presuming a universal scale of acceleration.  However, it should be pointed out that different proposed interpolating functions of MOND have all resulted in nonlinear  models   \citep{2012LRR....15...10F} while Eq. \ref{Potential} is clearly linear. This feature makes  analytical computations much easier as we will see in following. The linearity of the governing equation of motion could also lead to more straightforward galactic simulations. This is the subject of a future work.
 
\subsection{Comparing $\rho_{c_1}$ and various dark matter profiles }
\label{rhoc1}
A definitive solution to the problem of  mass discrepancy, whether it is found to be  dark matter or modified models of dynamics/gravity, has to explain the apparent successes of the rival models. For example, if models of dark matter are finally approved, they should be able to explain why the universal acceleration of MOND, and its   interpolating function, are valuable in explaining galactic systematic. On the other hand, a good modified model of gravity should illustrate the reason that some dark matter profiles are more consistent with the rotation curve   data. Here, we will explain a connection between hypothetical dark matter profiles and the extra term in the modified Poisson equation \ref{secondorder}.

\begin{figure}
\centering
\includegraphics[width=8cm]{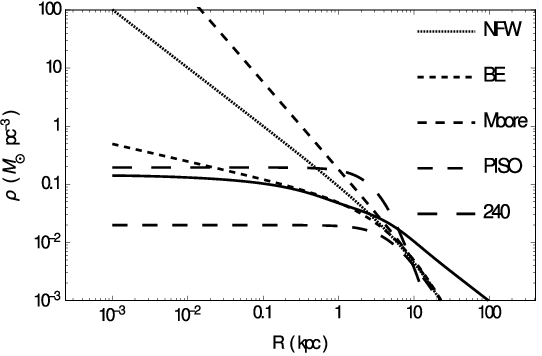}
\vspace{-12pt}
\caption{Comparing the prediction of $ \rho_{c_1} $ ( solid curve )  for the missing mass   with various dark matter density profiles of the Galaxy. Properties of different halos are derived from    \cite{2010A&A...509A..25W}. The CDM profiles are as follows: NFW \citep{1996ApJ...462..563N,1997ApJ...490..493N}, BE \citep{2001MNRAS.327L..27B}, Moore \citep{1999ApJ...524L..19M} and  pseudo-isothermal ( PISO ). In addition, we have plotted a halo profile named as 240 which is similar to the pseudo-isothermal model of CDM, though, it declines faster at outer parts of the galaxy compared to PISO \citep{2010A&A...509A..25W}.  \label{fig:rho}} 
\end{figure}

It is clear that the modified Poisson equation  \ref{secondorder} could be rewritten as $\nabla^{2}\Phi = 4 \pi G  (\rho_{m}+\rho_{c_1}) $
in which 
\begin{eqnarray}  \label{dens1}
\rho_{c_1} &=& \frac{c_{1}  a_0}{\pi G M} \int \frac{\rho_{m}(\vec{x}^{~\prime})d^{3}\vec{x}^{~\prime}}{| \vec{x}^{~\prime} - \vec{x}|}
\end{eqnarray}
plays the role of dark matter halo, i.e. it provides the system with more attractive force. In Fig. \ref{fig:rho},  we have plotted  various density profiles of CDM, for our Milky Way Galaxy, to make a comparison with $\rho_{c_1}$. The CDM profiles used here are Navarro, Frenk $\&$ White ( NFW ) \citep{1996ApJ...462..563N,1997ApJ...490..493N}, Binney $\&$ Evans ( BE )  \citep{2001MNRAS.327L..27B}, Moore \citep{1999ApJ...524L..19M} and  pseudo-isothermal ( PISO ). Also, the halo profile named as 240   is similar to the pseudo-isothermal model of CDM, though, it decreases faster at outer parts of the galaxy \citep{2010A&A...509A..25W}. The free parameters of these halos for the  Milky Way Galaxy are provided in Table. 1 of   \cite{2010A&A...509A..25W}.  The appropriate density and scale of the baryonic matter are also derived from    \cite{2010A&A...509A..25W}. As it is shown in Fig. \ref{fig:rho}, in very small radii the prediction of $\rho_{c_1}$ for the missing mass density is very close to the profile 240 of   \cite{2010A&A...509A..25W}; then, it converges mostly to BE profile for $0.1~ (kpc)~< R <~8~ (kpc)$. The density $\rho_{c_{1}}$ surely does not diverge at the center as some dark matter profiles do. See, for example, the behavior of NFW at $R \approx 0$ in Fig. \ref{fig:rho}. At large radii, $\rho_{c_1}$ shows a larger value compared to other profiles. This means a larger centripetal force compared to other dark matter halos at large radii. Of course, this results is obtained based on the assumption that the source of gravity is isolated.   However, most galaxies are within their group; thus, in general, one expects non-negligible external forces. The external forces might have some effects on galactic scaling rules as we will discuss below.

\subsection{Baryonic surface density and the dynamical acceleration}

 Assume the second-order modified Poisson equation of \ref{secondorder}. 
Since $\nabla^{2} \Phi =-\vec{\nabla}.\vec{a}$,  one could estimate  to the lowest order  the magnitude of    change in the gravitational acceleration due to an extended object  by $ |\nabla^{2} \Phi | \approx |\vec{a}|/L$, in which $ L$ is the typical size of the system. In addition, one may write $   \rho \propto  M/L^{3}$ and $   \int \frac{\rho(\vec{x}^{~\prime})d^{3}\vec{x}^{~\prime}}{| \vec{x}^{~\prime} - \vec{x}|} \propto  M/L$. Thus, one would expect a  relation between the magnitude of the centripetal acceleration,   and the baryonic surface density $\Sigma_{b} \propto M/L^{2} $  of the form of
\begin{eqnarray} 
 a=A_{MOD} \Sigma_{b} +B_{MOD}
\end{eqnarray}
in which $A_{MOD}$ and $B_{MOD}$ are some constants which should be determined from observations.
 Of course,  other modified  models of gravity are expected to display  somehow   similar relations at galactic scale, because, these models too are solely dependent to baryonic mass distribution and their scale  length. In the case of MOND, for example, we expect a relation as $a=A_{MOND} \Sigma^{1/2}_{b}$ ( see    \citep{2012LRR....15...10F} ) while for pure Newtonian theory, without dark matter halo, one could obtain a relation as  $a=A_{Newt} \Sigma_{b}$. The Poisson equation within the dark matter models, on the other hand, provides a relation between baryonic and halo mass and their appropriate scales. Therefore, a simple relation between the acceleration and the baryonic surface density is not expected through dark matter paradigm unless one could prove that there is a fine-tuned relation between baryonic and dark masses and disk-halo sizes.  To check the relation between the baryonic surface density and acceleration   we will use data from     \cite{2012LRR....15...10F} who provide the acceleration at the peak of the rotation curve $a_{p}$ and the characteristic surface density of the baryons.

\begin{figure}
\centering
\includegraphics[width=8cm]{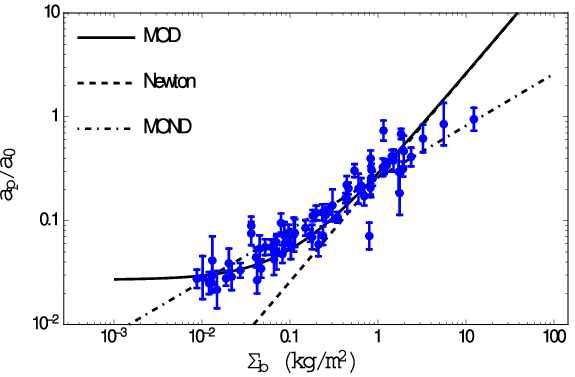}    
\vspace{-12pt}
  \caption{Typical  acceleration of galaxies ( in units of $a_{0} = 6.59 \times 10^{-10}~ m/s^{2}$ )  as a function of baryonic mass surface density $\Sigma_b ~$ ( in units of $ ~kg / m^{2}$ ). The acceleration $a_{p}$ is measured at the distance where the rotation velocity of baryons is maximum. Data is derived from     \cite{2012LRR....15...10F}. See also Fig.  7 of    \cite{2017ApJ...836..152L} for a similar graph which includes many individual measurements across galaxy rotation curves rather than using peak velocity for each object.}
  \label{fig:accsigma}
\end{figure}

In Fig. \ref{fig:accsigma}, we have plotted the data with theoretical curves of MOD, MOND and Newtonian models. The curves correspond to $A_{MOD}/a_{0} =0.26 m^{2}/kg$ and $B_{MOD}/a_{0}=0.027$ for MOD model while $A_{MOND}/a_{0} =0.26 m/kg^{0.5}$ and $A_{Newt}/a_{0}=0.26 m^{2}/kg $ for MOND  and Newtonian models,  respectively. In this plot, we have assumed $a_{0} =6.59 \times 10^{-10}~ m/s^{2}$ as mentioned above. As one could see from Fig. \ref{fig:accsigma}, the curves corresponding to MOND and MOD  are   more consistent with  the data.
 Although, in Fig. \ref{fig:accsigma} there appears a noticeable deviation  between the data and MOD at the very high surface densities, i.e. the last three data points on the right. However, even at this limit, the MOD curve is touched by the error bars of the two data points while the last data point displays a significant deviation  from the model. At low surface densities, on the other hand, the MOD curve follows the data pretty  well while MOND does the same two.
 The pure Newtonian line shows a significant discrepancy, especially for systems with lower surface densities. This is of course expected because the Newtonian model is not supposed to justify the galactic systematics without a dark matter halo. From this perspective, Fig. \ref{fig:accsigma} could be thought as an illustration for the mass discrepancy problem in Newtonian model. Also, it is seen from Fig. \ref{fig:accsigma} that the predictions of MOND and MOD are quite different for galaxies with very high and very low surface densities. Especially, the MOD model predicts a constant acceleration ( at  redshifts near zero ) for galaxies with very small densities while MOND predicts a decreasing acceleration as $a \propto \Sigma^{1/2}_{b}$. Thus, systems  with very low surface densities  could help to compare these two  models of modified dynamics.

Another important feature of MOND and MOD is that both these models predict an upper bound on $\Sigma$ \citep{1989ApJ...338..121M,shenavar2018local}. In fact,   \cite{1996MNRAS.280..337M} provides detailed analysis on disk galaxies and shows that the number of HSB galaxies with surface densities larger than $\Sigma_{\dagger} \equiv a_{0}/G$ decreases exponentially. This is while  HSB galaxies could be observed  more easily  because they have  higher surface brightness compared to LSBs. See, also, Fig. 6 in    \cite{shenavar2018local}. In Fig. \ref{fig:accsigma}, and  for the objects with very high surface densities, there is a clear deviation  between MOND and MOD  which could help us to compare the two models. More data on this tail of the plot could be helpful to decide between the two modified models of dynamics.

A similar graph to Fig. \ref{fig:accsigma}   could be seen in Fig. 7 of    \cite{2017ApJ...836..152L} , which is plotted based on many individual measurements of rotation curves. These authors investigate the link between baryons and dark matter in 240 galaxies and report that the observed acceleration  correlates ( over 4 dex ) with the expectations from baryonic  distribution. The relations is known as Radial Acceleration Relation (RAR) and it indicates a systematic deviation from Newtonian gravity  below the critical acceleration of $\sim 10^{-10}m/s^{2}$.  See also     \cite{2016PhRvL.117t1101M}.

 It is important to note that the discussion presented in this section is based on dimensional analysis; thus, Fig. \ref{fig:accsigma} essentially displays an estimated relation between a characteristic acceleration, i.e. $a_{p}$, and the baryonic surface density. For more accurate investigations, one should refer to accurate data analysis of galactic rotation curves or try to find more accurate "scaling rules",  as presented in the next sections.

\section{Galactic Scaling Rules Derived From the Virial Theorem}
\label{GSRV}
In this section, we will derive  four scaling relations based on the virial theorem. The validity of these scaling relations would be tested against SPARC data \citep{2016AJ....152..157L} in the following two sections by introducing  characteristic measures of velocity and size for  galactic systems.

Here, we will use the steady state  plus stability conditions  to derive the scaling rules of  galactic systems. If a system is in steady state, then its moment of inertia, i.e. $I$, is a constant. Using the scalar virial theorem,
\begin{eqnarray} \label{Virth}
\frac{1}{2}\frac{d^{2}I}{dt^{2}}=2T+W,
\end{eqnarray}
in which $T \equiv 1/2 \int \rho d^{3}\vec{x}~  \vec{v}^{2} $ and $W \equiv -\int  \rho d^{3}\vec{x}~ (\vec{x}.\vec{\nabla} \Phi)$ stand for the total kinetic energy and the virial energy respectively, one could write the steady state condition as 
\begin{eqnarray}  \label{Virial}
2T+W =0.
\end{eqnarray}
See   \cite{1978vtsa.book.....C} and also chapters 4 and 7 of   \cite{2008gady.book.....B} for more details.

\begin{table}   
\centering  
  \begin{tabular}{|c|c|c|}   
    \hline
Geometry     & $\alpha$ & $\beta$ \\
\hline
Homogeneous sphere   & 3/5 & 18/35  \\
Exponential sphere   & 5/32 & 7/4   \\
Exponential disk (thin)     & $11.63/(4\pi^{2})$ & 1.47 \\
Exponential disk (thick)     & 0.28 & 1.48 \\
\hline
\end{tabular}
\vspace{-7pt}
\caption{The virial parameters $\alpha$ and $\beta$. The virial parameters for the first three geometries are derived by    \cite{Shenavar:2016xcp}. By the same method, one could derive the virial parameters for a thick exponential disk, i.e. the last case. }
\label{table:paras} 
\end{table}

For a single-component gravitating system, and by using our modified dynamics, it has been previously proved that the virial energy is of the general form of 
\begin{eqnarray} \label{Virenergy}
W(R) = -\alpha \frac{GM^{2}}{R} - 2c_{1}\beta Ma_{0}R
\end{eqnarray}  
in which $\alpha$ and $\beta$ are some positive parameters solely dependent to the mass distribution of the system \citep{Shenavar:2016xcp}. See Table \ref{table:paras} for typical values of $\alpha $ and $\beta$. The virial parameters   $\alpha $ and $\beta$ for the first three geometries are derived in Appendix of   \citep{Shenavar:2016xcp}.   For the last case, i.e. thick exponential disk, the parameters $\alpha $ and $\beta$ are derived numerically by assuming that the vertical mass distribution of the disk changes as $sech(z/z_{0}) /(\pi z_{0}) $ in which $z_{0}$ is the scale height of the system presumed to be $1/10$ of the disk radius. In fact, one could show that by assuming the  vertical mass distribution  as $sech^{2}(z/z_{0})/(2 z_{0}) $, the results for $\alpha$ and $\beta$ would be almost the same. Thus,  $\alpha = 0.28$ and $ \beta = 1.48$ appropriately represent the virial parameters of a thick exponential disk.

\begin{figure}
\centering
\includegraphics[width=8cm]{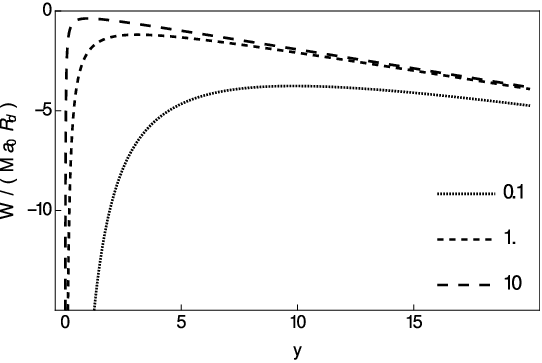}
\vspace{-15pt}
\caption{The virial energy of a thin exponential  disk ( in units of  $Ma_{0}R_{d}$ ) as a function of $y=R/(2R_{d})$ and for different values of Freeman ratio $\mathcal{R}_{F} =0.1,~1.0,~10 $. To plot this graph, we have assumed $ \alpha = 11.63/(4\pi^{2})$ and $ \beta = 1.47$. It is easy to see that every system possesses a maximum of virial energy which corresponds to a minimum of kinetic energy. In addition, for systems with smaller Freeman ratio ( denser galaxies ), the maximum occurs in larger radii ( compared to $R_{d}$ ).   \label{fig:VirialPlot}} 
\end{figure}

In addition to the steady state condition, i.e. Eq. \ref{Virial}, one needs a stability criterion to explain the behavior of the gravitating systems. Here, we maximize the virial energy to find the stable state. The idea of maximizing virial energy to find the scaling relation of galactic systems is first exploited by Secco. See, for example, \cite{2000NewA....5..403S, 2001NewA....6..339S}. From physical point of view, Eq.  \ref{Virial} suggests that  when the virial energy is maximum, the kinetic energy  should be minimum. See Fig. \ref{fig:VirialPlot} which displays the virial energy of a thin exponential  disk ( with the scale length $R_{d}$  ) as a function of $y=R/(2R_{d})$   for different values of Freeman ratio $\mathcal{R}_{F} \equiv   \Sigma_{\dagger} / \Sigma_{0} =0.1,~1.0,~10 $.  We have chosen the thin exponential disk for the sake of convenience; however, it could be easily checked that the general behavior of $W(R)$ in Eq. \ref{Virenergy}  is  similar for other mass distributions.

 It is clear from Fig. \ref{fig:VirialPlot}  that for systems with smaller Freeman ratio, i.e. denser systems, the maximum of the virial function occurs in larger radii. In fact, it could be proved that for any thin exponential  disk one has $ y_{max} \propto \mathcal{R}^{-1/2}_{F}  $ in which $y_{max}$ is the radius for which the maximum of $W(R)$ happens. Thus, the maximum point of denser systems occurs in larger radii ( in units of $2R_{d}$ ). Moreover, the maximum point of  systems with smaller Freeman ratio occurs in smaller virial energies, i.e. larger values of $|W(R)|$. Thus, from Eq. \ref{Virial}, one could immediately realize that because in the steady state condition we have $T=-1/2W$, the dense systems possess larger kinetic energies, i.e. greater velocities for denser galaxies. This behavior has been reported before based on the observations of galaxies \citep{1991AJ....101.1231C}.

Now, we obtain the four scaling rules by assuming an stable equilibrium for galactic systems. The derivation is quite general because we only presume the virial theorem \ref{Virial} and the general form of the virial energy \ref{Virenergy}. Hence, no specific mass distribution is assumed. To do so, we note that for a general system with the virial energy of   \ref{Virenergy}, and by maximizing  $W(R)$ function,  one could find the radius in which the system is in stable virial equilibrium as 
\begin{eqnarray}   \label{RM}
R_{eq}=\sqrt{\frac{GM\alpha}{2c_{1}a_{0}\beta}}.
\end{eqnarray}
This is one of the scaling rules.  This relation is reminiscent of the definition of the MOND transition radius for a point mass.  It is easy to see that, except for the constants of the model like $a_{0}$ and $\beta$ etc., this expression for the equilibrium radius is only related to the mass of the system.

Now, using the steady state condition  \ref{Virial}  and the stability condition    \ref{RM}  one could find a relation between the total mass of the system and its characteristic velocity in virial equilibrium $V_{eq}$ as 
\begin{eqnarray}  \label{TF}
V^{4}_{eq}=8c_{1}a_{0}\alpha \beta GM
\end{eqnarray}
 To derive this relation, we replace $T=1/2MV^{2}_{eq}$ and $W|_{R_{eq}}$ from Eq. \ref{Virenergy} into Eq. \ref{Virial}. This is the second scaling relation (BTFR).

The scale length of the system could also be found by eliminating mass $M$ from Eqs. \ref{RM} and \ref{TF} as
\begin{eqnarray} \label{RV}
R_{eq}=\frac{V^{2}_{eq}}{4\beta c_{1} a_{0}}.
\end{eqnarray}
which, except for the constants of the model,  is only dependent to the virial velocity $V_{eq}$ of the system. Equation \ref{RV} is the third scaling relation. It is also possible to derive this scale length directly from the steady state condition of \ref{Virial} 
\begin{eqnarray}  \nonumber
-\alpha \frac{GM^{2}}{R_{eq}} - 2c_{1}\beta Ma_{0}R_{eq} + MV^{2}_{eq}=0
\end{eqnarray}
where, by solving for $R_{eq}$ results in
\begin{eqnarray}  \nonumber
R_{eq}=\frac{V^{2}_{eq} \pm \sqrt{V^{4}_{eq}-8c_{1}a_{0}\alpha \beta GM}}{4\beta c_{1} a_{0}}.
\end{eqnarray}
Using Eq. \ref{TF} we see that in the last equation we have $V^{4}_{eq}-8c_{1}a_{0}\alpha \beta GM=0$ and therefore we obtain Eq. \ref{RV} again. 

Finally, combining any pair of these three scaling relations, i.e. \ref{RM}, \ref{TF} and \ref{RV}, one can find another scaling law between mass, velocity and scale length as:
\begin{eqnarray} \label{RVM}
V^{2}_{eq} R_{eq}=2\alpha GM.
\end{eqnarray}
These four scaling rules are closely interconnected because they have been derived from simultaneous imposition of stability and equilibrium conditions in the context of modified dynamics \ref{Potential}. However, only two of the relations could be considered as independent as discussed in the Introduction.
 Being derived from the virial theorem, these scaling relations characterize the whole system and in particular they do not hold at a certain radial distance.

\begin{figure}
\begin{center}
\begin{tikzpicture}
\draw (0,0) rectangle (3,1);
\draw (4.5,0) rectangle (7.5,1);
\draw (0,-2) rectangle (3,-1);
\draw (4.5,-2) rectangle (7.5,-1);
\node at (1.5,.5) {$R_{eq}=\frac{V^{2}_{eq}}{4\beta c_{1} a_{0}}$};
\node at (6,.5) {$R_{eq}=\sqrt{\frac{GM\alpha}{2c_{1}a_{0}\beta}}$};
\node at (1.5,-1.5) {$V^{4}_{eq}=8c_{1}a_{0}\alpha \beta GM$};
\node at (6,-1.5) {$V^{2}_{eq} R_{eq}=2\alpha GM$};
\draw [decoration={
markings,mark=at position 1cm with {\arrow{Latex}}},postaction={decorate}](1.3,0) -- (1.3,-1);
\draw [decoration={
markings,mark=at position 1cm with {\arrow{Latex}}},postaction={decorate}] (1.6,-1) -- (1.6,0);
\draw [decoration={
markings,mark=at position 1cm with {\arrow{Latex}}},postaction={decorate}](3,0.6) -- (4.5,0.6);
\draw [decoration={
markings,mark=at position 1cm with {\arrow{Latex}}},postaction={decorate}](4.5,0.4) -- (3,0.4);
\draw [decoration={
markings,mark=at position 1cm with {\arrow{Latex}}},postaction={decorate}](5.9,0) -- (5.9,-1);
\draw [decoration={
markings,mark=at position 1cm with {\arrow{Latex}}},postaction={decorate}] (6.2,-1) -- (6.2,0);
\draw [decoration={
markings,mark=at position 1cm with {\arrow{Latex}}},postaction={decorate}](3,-1.6) -- (4.5,-1.6);
\draw [decoration={
markings,mark=at position 1cm with {\arrow{Latex}}},postaction={decorate}](4.5,-1.4) -- (3,-1.4);
\end{tikzpicture}
\end{center}
\vspace{-18pt}
\caption{The Four scaling relations which are derived  from virial theorem plus assuming a new modified dynamics governed by potential \ref{Potential}. The parameters $\alpha$ and $\beta$ are dimensionless factors  of order unity which are reported in Table. \ref{table:paras} for different geometries.}
  \label{fig:scaling}
\end{figure}
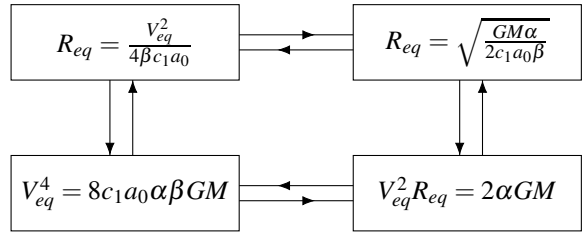

The interconnectivity of these four scaling rules is displayed in Fig. \ref{fig:scaling}. As   \cite{2017ApJ...836..151S} has explained, versions of these relations have been well known in the literature \citep{1960ApJ...131..293B,1981MNRAS.194..809L,1995AIPC..336..495K}; though, in some prior works, researchers have used magnitude rather than baryonic mass to discuss the relations.  Schulz  derives all of these scaling rules by an elegant line of reasoning based on two propositions. The first assumption is that  the centripetal acceleration at the edge of the galaxy is proportional to the predicted acceleration  of Newtonian physics. Second, this acceleration is a constant which is related to MOND constant acceleration  $a_{0MOND}$.   However, it should be mentioned that there is another constant in Schulz's proposal which is named as flatness factor or $C_{f}$. This constant is defined as the "increased gravitational force due to a flattened mass distribution versus a point mass" \citep{2017ApJ...836..151S}. The constant $C_{f}$ helps to fit the data to the scaling rules; though, it only changes the intercepts in $\log - \log$ planes. See our discussion on data analysis below for a similar situation.

While Eq. \ref{RVM} appears usually in researches  related to the dark matter paradigm (it is also the oldest of the four rules \citep{1960ApJ...131..293B} ),  some scaling relations similar to  Eqs. \ref{RM}, \ref{TF} and \ref{RV} have attracted much attention in MOND literature.  In fact, by  noticing that the typical centripetal acceleration at the edge of  galactic systems is almost constant, i.e. $V^{2}/R \approx a_{0}$,   \cite{Milgrom:1983ca} started the whole MOND model. This observation is similar to Eq. \ref{RV}. Moreover, Milgrom explained the mass-velocity scaling rule, i.e. $M \propto V^{4}$, which was first reported by   \cite{1977A&A....54..661T} as a relation between luminosity and circular velocity.

In addition to the above method, it is also possible to obtain the four scaling relations of Fig. \ref{fig:scaling} by dimensional analysis of the modified Poisson equation. We have reported such derivation in  \ref{Derive}. In comparison, however, the virial method manifests its advantage over the method of dimensional analysis because by using the former method one could introduce  measures of virial size and velocity of the systems ( and their uncertainties ). Then, it is possible to employ these measures to investigate the reliability of the scaling rules using relatively accurate data. In what follows, we introduce  measures of galactic mass, size and velocity for the  SPARC sample, and subsequently, we will study the slopes and intercepts of the scaling relations.

\section{Measures of Galactic Mass, Size and Velocity in Virial Equilibrium}
\label{Measures}
In this section, we will define proper measures of mass, size and velocity of galactic systems based on data provided by SPARC \citep{2016AJ....152..157L}. 
 It should be mentioned that these three parameters are    introduced in a way  that they represent the whole system ( stellar + gas ).

\subsection{SPARC data}
\label{SPARC}
The SPARC (Spitzer Photometry and Accurate Rotation Curves)  galactic sample, provided by   \cite{2016AJ....152..157L}, includes 175 nearby objects with high-quality data for rotation velocity derived based on HI/H$\alpha$ studies.  SPARC also uses new surface photometry at  $ 3.6~ \mu m$  which traces the stellar mass.   \cite{2016ApJ...816L..14L} employ  SPARC data to study the  BTFR. To do so, they provide an algorithm to exclude galaxies with rising rotation velocity $V_{f}$, some systems with very high inclination corrections and also objects with low-quality rotation curves. This leaves the sample with 118 galaxies. 

 Four objects ( out of 118 ) do not possess the scale length of the stellar disk $ R_{d} $ and/or  $R_{HI}$. The quantity $R_{HI}$ is the radius where the $HI$ surface density, corrected to face-on, reaches $1~M_{\odot}/pc$. These two scales  are needed  ( as we will see below )  to define a suitable virial size for galaxies. Thus, these systems, i.e. NGC 5907, D 631-7, NGC 5533 and NGC 7339, will be excluded from our analysis.   Therefore, our sub-sample includes 114 objects. This sub-sample  of SPARC is suitable for our data analysis because it covers a wide range of galactic morphology ( S0 to Im ), galactic radius $(~ 0.3~ kpc \lesssim  R \lesssim 15~ kpc ~)$,  gas fraction $f_{g}=M_{g}/M_{b}$ of the systems $(~ 0.01  \lesssim  f_{g} \lesssim 0.95 ~)$, galactic baryonic mass  $(~10^{8} M_{\odot} \lesssim  M_{b} \lesssim 10^{11} M_{\odot} ~)$ and surface brightness of the systems  $(~8 ~  L_{\odot}/pc^{2} \lesssim  \Sigma \lesssim 3000  ~ L_{\odot}/pc^{2}~)$. In addition,  merging systems have been excluded from the main sample. This point makes the data even more appropriate since our arguments are essentially based on virial theorem which becomes significantly more complicated in the case of merging systems. Beside this, merging systems display  out-of-equilibrium HI kinematics which  artificially increase  the final scatter in the results. In conclusion, the sample which is used here represents  non-merging objects in environments with small densities.

\subsection{The measure of mass} 
Virial method dictates the use of total baryonic mass  instead of luminosity, and in our following data analysis, we will employ this measure of mass to carry out the calculations. In fact,    \cite{2000ApJ...533L..99M} have shown that  the optical Tully-Fisher relation breaks down for velocities  less than $  100~ km / s$ while by considering the total baryonic mass, the  relation is tight over five decades of stellar mass and circular velocities ranging over $30~ km/ s \lesssim  V \lesssim 300~ km/ s$.  As  \cite{2016ApJ...816L..14L} have explained ( Sec. 2.3 ) the total baryonic mass  in SPARC is derived from $M_{b}= M_{g}+\Upsilon_{\star}L_{[3.6]}$ in which $M_{g} =1.33M_{HI}$ is the mass of gass, 1.33 being included to contain the share of helium, $\Upsilon_{\star}$ is the mass-to-light ratio of stars  assumed to be $0.5$ throughout and finally $L_{[3.6]}$ is the $[3.6]$ luminosity. The uncertainty in $M_{g}$ is reported to be generally smaller than $10~\%$ which will be used to estimate $\delta f_{g}/f_{g}$ in Eq. \ref{errorfg} below. The collected uncertainty in the total baryonic mass $M_{b}$ due to errors in $M_{g}$, $L_{[3.6]}$, $\Upsilon_{\star}$ and distance has been carefully calculated by   \cite{2016ApJ...816L..14L}. Moreover, the error in flat velocity $V_{f}$ due to non-rotational velocities, asymmetries between the two sides of the disk, the dispersion around average velocity and  the inclination correction is also provided in SPARC.

\subsection{Equilibrium radius of real systems}
To fit the above scaling relations, one needs to introduce an appropriate dynamical length scale for real galactic systems. This length scale must represent the two components of the systems, i.e. stellar and gas   ( hereafter shown by indices $s$ and $g$ respectively ), and their collective gravitational effects. However, this is not a trivial work due to some observational and theoretical reasons. From observational point of view, the issue is that one can define many scale lengths for the stellar component such as the effective radius $R_{eff}$, which  encompasses half of the total luminosity, and also $R_{d}$ which provides  the scale length of the stellar disk. The scales $R_{eff}$ and $R_{d}$  are provided in SPARC catalog. On the other hand, for the gaseous component, SPARC provides $R_{HI}$ as defined above. None of these scale length are necessarily   compatible with the statistical nature of the virial theorem; however, these are the best available options provided by SPARC.  There are other characteristic radii such as $R_{200}$, which is defined as the length where the average  density is $200$ times larger  than the critical density  of the universe. We do not consider this scale length here.

From theoretical point of view,  the main issue   is to find  a dynamical scale length, i.e. a combination of  $R_{d}$ ( or $R_{eff}$ ) and $R_{HI}$,  to represent the virial size of the systems. To do so, it is necessary to derive a radius based on virial energy  of compound ( gas + stars ) systems. We have provided the detailed calculations of the virial energy of a system consisting of gas and stars in   \ref{Comound}. To first order of approximation, the virial energy of such system is 
\begin{eqnarray}  \label{Wtot}
-W_{tot} &\simeq &  G \left[ \alpha (  \frac{M^{2}_{g}}{R_{g}} + \frac{M^{2}_{s}}{R_{s}}) + \gamma_{1} \frac{M_{s}M_{g}}{R_{g}}  \right]  + \frac{2 c_{1}a_{0}  }{M}  \left[ \beta ( M^{2}_{g} R_{g} + M^{2}_{s} R_{s} ) +\gamma_{2} M_{s}M_{g} R_{g}  \right] 
\end{eqnarray}
in which $\gamma_{1}$ and $\gamma_{2}$ are some dimensionless parameters related to the gravitational interaction of gaseous and stellar components. Also, $M_{s} $ $(~ M_{g} ~)$ and $R_{s} $ $(~ R_{g} ~) $ are the mass and scale length of the stellar ( gaseous ) system respectively  while $M$ is the total mass of the system. In the above formula, we have presumed that stellar and gas systems have the same geometry; thus, they both have the same virial coefficients $\alpha$ and $\beta$. This simplification reduces the number of degrees of freedom in our data analysis.

Now,  to find the equilibrium radius, imagine that there is  a system with total mass $M$,  radius $R$ and  the  virial energy of
$$ -W = \alpha^{\prime} \frac{GM^{2}}{R} + 2c_{1}\beta^{\prime} Ma_{0}R  $$
which is equivalent to the virial energy of of the original system, i.e. $W = W_{tot}$. By equating the last two formulas for virial energy, one could find a quadratic  equation for $R$ with two solutions. We define the arithmetic mean of the two roots as the equilibrium radius: 
\begin{eqnarray}  
R_{eq}= \frac{ GM}{4 c_{1}a_{0} \beta^{\prime}} \left[ \alpha( \frac{f^{2}_{g}}{R_{g}} + \frac{(1-f_{g})^{2}}{R_{s}}) + \gamma_{1}  \frac{f_{g}(1-f_{g})}{R_{g}}  \right]  + \frac{1 }{2\beta^{\prime}}  \left[ \beta ( f^{2}_{g} R_{g} + (1-f_{g})^{2} R_{s} ) + 
\gamma_{2} f_{g} (1-f_{g}) R_{g } \right] 
\end{eqnarray}
in which $f_{g}=M_{g}/M$ is the ratio of the mass of the gas to the total baryonic mass. There are six dimensionless parameters, i.e. $c_{1}$, $\alpha$,   $\beta$,  $\beta^{\prime} $, $\gamma_{1}$ and $\gamma_{2}$,  in the definition of $ R_{eq} $ which only four of them could be independent. Thus, we redefine these parameters as $\lambda_{1}=\alpha/ (4\beta^{\prime} c_{1})$, 
$\lambda_{2}=\gamma_{1} / (4\beta^{\prime} c_{1})$, $\lambda_{3}=\beta/ (2\beta^{\prime})$, $\lambda_{4}=\gamma_{2} / (2\beta^{\prime})$ to simplify the expression for $ R_{eq} $ as
\begin{eqnarray} \label{radius}
R_{eq}=  \frac{ GM}{ a_{0} } \left[ \lambda_{1}( \frac{f^{2}_{g}}{R_{g}} + \frac{(1-f_{g})^{2}}{R_{s}}) + \lambda_{2} \frac{f_{g}(1-f_{g})}{R_{g}} \right]  +   \lambda_{3} ( f^{2}_{g} R_{g} + (1-f_{g})^{2} R_{s} ) + \lambda_{4} f_{g} (1-f_{g}) R_{g } 
\end{eqnarray}
Also, in our data analysis below, we have used the value  $\Sigma_{\dagger} = a_{0}/G = 4.7 \times 10^{9} M_{\odot}/kpc^{2}$ for the critical surface density. 

\begin{figure}
\begin{center}
\includegraphics[width=8.5cm]{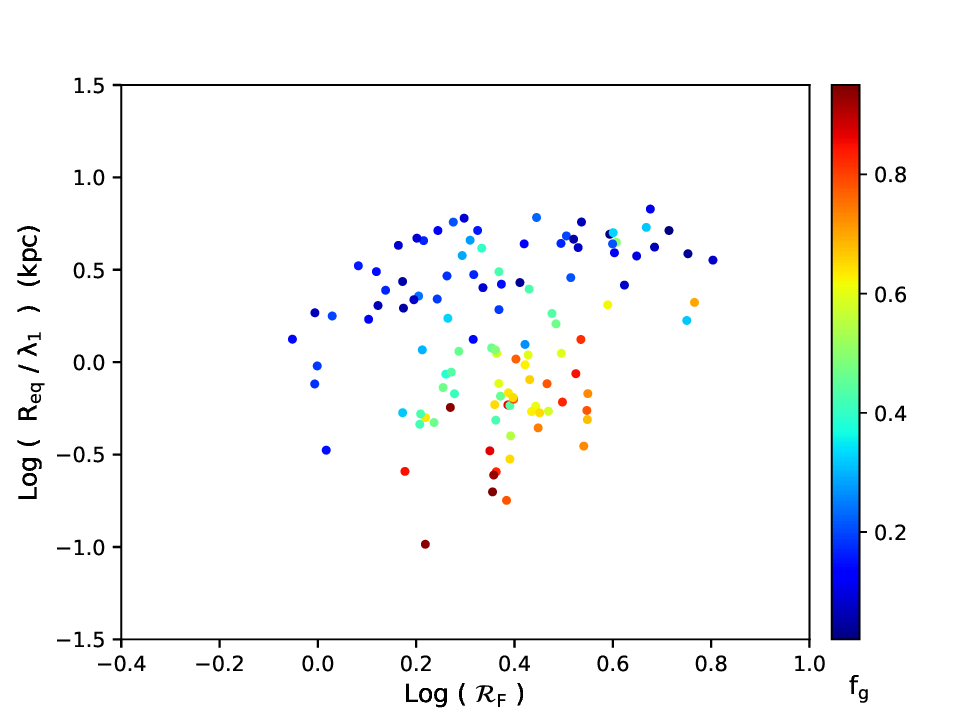}
\vspace{-15pt}
\caption{ Equilibrium radius $R_{eq}$ as a function of  Freeman ratio $\mathcal{R}_{F}$. In this plot it is assumed that $R_{s}=R_{eff}$ and $(~ \lambda_{2},~ \lambda_{3}, ~ \lambda_{4}~)/\lambda_{1} = (~1.1,~0.039,~0.19~)$. It could be seen that the gas-rich systems ( galaxies with high $f_{g}=M_{g}/M_{b}$ ) are separated from gas-poor ones. A similar graph could be achieved if one assumes $R_{s}=R_{d}$ in deriving $R_{eq}$. See also Fig. \ref{fig:Freeman} which shows the change in $R_{HI}$, $R_{eff}$ and $R_{d}$ as a function of  Freeman ratio. Color online.  \label{fig:FreemanReq}}
\end{center}
\end{figure}

To see the behavior of the equilibrium radius as a function of Freeman ratio $ \mathcal{R}_{F} = \Sigma_{\dagger} /\Sigma_{0} $,  we have plotted $R_{eq}$  in Fig. \ref{fig:FreemanReq} for the objects of  SPARC sample. Here,  $\Sigma_{0} =M/\pi R^{2}_{eq}$ is assumed as the proper surface density of the system with the scale length $R_{eq}$. Also, the plot is drawn by assuming $R_{s}=R_{eff}$ and $(~ \lambda_{2},~ \lambda_{3}, ~ \lambda_{4}~)/\lambda_{1} = (~1.1,~0.039,~0.19~)$ which are the best parameters derived from the data analysis in Sec. \ref{Datan}.  A similar graph could be produced if one assumes $R_{s}=R_{d}$ in deriving $R_{eq}$. See also Fig. \ref{fig:Freeman} which shows the change in $R_{HI}$, $R_{eff}$ and $R_{d}$ as a function of  Freeman ratio. By comparing Figs. \ref{fig:FreemanReq} and \ref{fig:Freeman}, it could be seen that the behavior of $R_{eq}$  as a function of $ \mathcal{R}_{F}$ is more similar to $R_{HI}$ than $R_{eff}$ or $R_{d}$.

 Eq. \ref{radius} means that  when $ \mathcal{R}_{F} $ is small, the equilibrium radius is determined by the first term on the RHS of this equation, i.e. the term proportional to $GM/a_{0}$. In other words, when $ \mathcal{R}_{F} $ is small, the system is dominated by the Newtonian gravitational energy. On the other hand, when $ \mathcal{R}_{F} $ is large, the terms proportional to $\lambda_{3}$ and $\lambda_{4}$ in  Eq. \ref{radius} have a dominant share  in the equilibrium radius. These terms are due to MOD virial energy. In conclusion, $R_{eq}$ puts galaxies with large and small Freeman ratios on the same plane; thus, the scatter in size-dependent  relations ( $M-R$ and $V-R$ ) decreases significantly. 
 
 Another advantage of using $R_{eq}$ as a scale length of galactic systems ( instead of $R_{d}$ or $R_{eff}$ ) is that $R_{eq}$  includes the stellar and gaseous sizes with their appropriate share in the systems ( $1-f_{g}$ and $f_{g}$ respectively). This is particularly important because in some systems, the gas contributes more mass than the stars ( up to 95 $\%$ in a few objects of SPARC  ). Of course, using $R_{d}$ or $R_{eff}$ as the representative of the scale of the system would not be fitting in such situation anymore.

 It could be seen in Fig. \ref{fig:FreemanReq} that the gas-rich systems are separated from gas-poor ones. Moreover, very few objects could be found with $ \mathcal{R}_{F} <1 $.  This is due to the reason that in MOD dynamics,  systems  with $\Sigma_{0}  \gtrsim  \Sigma_{\dagger}$ are locally unstable. See     \cite{shenavar2018local} for the detailed proof.

The equilibrium radius $R_{eq}$ depends on four observable parameters   $R_{s}$, $R_{g}$, $M$ and $ f_{g}$. The error in this radius could be estimated as 
\begin{eqnarray} \label{error}
\delta R_{eq} =  \left( ( \frac{\partial R_{eq}}{\partial R_{s}} \delta R_{s}  )^{2} + ( \frac{\partial R_{eq}}{\partial R_{g}} \delta R_{g}  )^{2} + ( \frac{\partial R_{eq}}{\partial M} \delta M  )^{2} + ( \frac{\partial R_{eq}}{\partial f_{g}} \delta f_{g} )^{2} \right)^{1/2}
\end{eqnarray} 
in which 
\begin{eqnarray} \label{errorfg}
 \frac{\delta f_{g}}{f_{g}} =   \sqrt{ \left( \frac{\delta M_{g}}{M_{g}} \right)^{2} + \left( \frac{\delta M}{M} \right)^{2} }
\end{eqnarray}
represents the error in gas fraction. All necessary quantities to calculate $ \delta R_{eq}$ are provided by SPARC.

\subsection{Equilibrium Velocity}
 The kinetic energy in virial theorem \ref{Virth} contains both rotational and random terms. In the same way, the equilibrium velocity $V_{eq}$, which represents a system in virial equilibrium, must contain both rotation and pressure tracers. A suitable velocity which fulfills this condition could be introduced as
\begin{eqnarray}
V_{eq} = \left( 0.5 V^{2}_{f} + \sigma^{2}_{eq}  \right)^{1/2}
\end{eqnarray}
where $\sigma_{eq}$ is the velocity dispersion, i.e. it is a measure of disordered or noncircular motions. Some authors \citep{2006ApJ...653.1027W,2007ApJ...660L..35K,2014ApJ...795L..37C}  have proposed a similar measure of velocity to include the share of the gas velocity dispersion. However, here $\sigma_{eq}$ represents the velocity dispersion in both gas and stars.

\begin{table}
\begin{center}

\begin{small}   
  \begin{tabular}{|c|c|c|c|c|c|c|c|c|c|c|c|c|c|c|c|c|}    
    \hline 
     \multicolumn{4}{|c|}{Fitting Parameters}& \multicolumn{4}{c|}{$\log R_{eq}=a+b(\log M_{b} - \log M_{b0})$} &\multicolumn{4}{c|}{$\log R_{eq}=a+b(\log V_{eq} - \log V_{eq0})$} & \multicolumn{4}{c|}{$\log M_{b}= a+b(\log V_{eq} - \log V_{eq0})$} &  \\  \hline
    $\frac{\lambda_{2}}{\lambda_{1}}$ & $\frac{\lambda_{3}}{\lambda_{1}}$  & $\frac{\lambda_{4}}{\lambda_{1}}$ & $\frac{\sigma_{eq}}{(km/s)} $ & $a$ & $b$& $\Delta$ & $\epsilon_{y}$  & $a$ & $b$& $\Delta$ & $\epsilon_{y}$   & $a$ & $b$& $\Delta$ & $\epsilon_{y}$  &   \\
    \hline
0.80 & 0.046 &	0.21 &	14 & 0.178 & 0.483 & 0.080 & 0.0 & 0.178 & 1.865 & 0.13 & 0.0 & 10.136 & 3.91 & 0.21 & 0.082   &  \\
0.95 &	0.043 &	0.17 &	13 &  0.127 & 0.497 & 0.083 & 0.0 & 0.134 & 1.937 & 0.13 & 0.0 & 10.137 & 3.889 & 0.21 & 0.082  &  \\    
 1.10 &	0.039 &	0.19  &	15 & 0.146 & 0.498 & 0.082 & 0.0 & 0.15 & 1.937 & 0.13 & 0.0 & 10.134 & 3.932 & 0.21 & 0.083  &  $ \surd$  \\   
1.10 & 0.036 & 0.18 & 10  & 0.136 & 0.503 & 0.071 & 0.0 & 0.137 & 1.933 & 0.13 & 0.0 & 10.142 & 3.834 & 0.21 & 0.081  & \\
1.30 & 0.037 &	0.18 &	14 & 0.140 & 0.502 & 0.071 & 0.0 & 0.138 & 1.957 & 0.13 & 0.0 & 10.136 & 3.910 & 0.21 & 0.082  &  \\
1.45 &	0.037 &	0.19 &	17  & 0.148 & 0.490 & 0.083 & 0.0 & 0.149 & 1.963 & 0.13 & 0.0 & 10.130 & 3.981 & 0.21 & 0.085  &  \\
1.55 &	0.035 &	0.21 &	11 & 0.169 & 0.487 & 0.083 & 0.0 & 0.163 & 1.870 & 0.13 & 0.0 & 10.141 & 3.851 & 0.21 & 0.081   &  \\ \hline
  \end{tabular}
  \vspace{-12pt}
  \caption{ Results from fitting the bivariate scaling relations of Fig. \ref{fig:scaling}. It is assumed that the size of the stellar disk is equal to $R_{eff}$. Data from SPARC \citep{2016ApJ...816L..14L,2016AJ....152..157L}. The plots corresponding to the set of parameters indicated by $ \surd $ are provided in Figs \ref{fig:R2M}, \ref{fig:V2R} and \ref{fig:TF}. } 
\label{table:Reff} 
\end{small}
\end{center}
\end{table}

  It could be easily seen that although the correction due to $\sigma_{eq}$ is negligible for most galaxies with $ V_{f} \gtrsim 40 $,  the overall effect of this term  is considerable. To see this point, consider BTFR.  In this relation, the correction of $\sigma_{eq}$ shifts the objects with smallest $V_{f}$ toward right on velocity axis while it keeps data points with high velocities almost unchanged. Thus, the slope in BTFR   increases from   $\simeq 3.7 $ to $\simeq 3.9 $ and higher. Compare Figs \ref{fig:TF} and \ref{fig:TFsig0} to see this point.  
  Adding    $\sigma_{eq}$, which we assume to be smaller than $20 ~km/s$ in our data analysis, could help us  to find unified scaling rules for  pressure-dominated  and  rotation-dominated  galaxies; though, this matter needs a much larger sample and is beyond the scope of our work.

 The error in $V_{eq}$ could be simply estimated because the error in  $ V_{f}$ is provided by SPARC. We should mention that  \cite{2019MNRAS.484.3267L} have reported  a minor indexing bug  in the calculations of the error in $V_{f}$ provided by \cite{2016ApJ...816L..14L}. However,  this bug causes an underestimation of the errors by $0.6 ~km~ s^{-1}$ on average  which corresponds to $0.1~ \%$ deviation in the final results; thus, we neglect it here.

\section{Data Analysis} 
\label{Datan} 
\subsection{Method of data analysis }
From statistical analysis point of view, and due to large uncertainties in most galactic data, the data analysis of galactic scaling rules is a challenging issue. The related least-square  problem has been discussed in many works  \citep{1996ApJ...470..706A,2002ApJ...574..740T,press2007numerical}. Here, we use the  method which has been explained in Sec. 3.2 of   \cite{2013MNRAS.432.1709C} to perform robust line   fit  to data with uncertainties in all parameters and also possible unknown intrinsic scatter\footnote{The LTS-LINEFIT code is available here: \url{https://www-astro.physics.ox.ac.uk/~mxc/software/}.}. This method combines the Least Trimmed Squares (LTS) robust procedure which is  suggested by   \cite{rousseeuw1984least}  for the first time and then improved by   \cite{rousseeuw2006computing}.  This technique  also forces the fit to converge to the appropriate solution in the presence of significant outlier data points \citep{1987rrod.book.....L}. The outlier data are shown by green diamonds in all fitting plots below. Cappellari's code  automatically excludes the outliers from the fit.

For a thorough review on the method which is used here, one could refer to \cite{2013MNRAS.432.1709C}. To fit a linear relation of the form $y=a+b(x-x_{0})$ between two data sets $x_{i}$ and $y_{i}$, with error bars on both axes $\delta x_{i}$ and $\delta y_{i}$, and also intrinsic scatter
in the y-coordinate $\epsilon_{y}$, one can  minimize the quantity
\begin{eqnarray}
\chi^{2} = \sum^{N}_{i} \frac{\left(a +b(x_{i}-x_{0})-y_{i} \right)^{2}}{(b\delta x_{i})^{2}+\delta y^{2}_{i}+\epsilon^{2}_{y}} 
\end{eqnarray}
in which N is the number of data points, to find the best fit. The value of the intrinsic scatter is found iteratively by forcing the fit toward the point that $\chi^{2}$ per degrees of freedom , i.e. $\chi^{2}/(N-2)$, converges to unity. In other words, in this method the chi-by-eye rule is always satisfied by enforcing it on the fit.

 In each of the following fits, the observed root-mean-square ( rms ) scatter $\Delta $ in dex would be provided in the graph.   In addition, the plot for Studentized residual is provided in every case. Also, in the residual plots  we have used Grubbs's critical value, as a simple measure,  to indicate the possible outliers  \citep{grubbs1969procedures}. 

The scaling relation $ M \propto V^{2}R $ could be analyzed by a  plane fit method which is also provided by  \cite{2013MNRAS.432.1709C}.  In this case, however, there appears the issue of collinearity of $V$ and $R$ parameters because, as we will see below, these two parameters are strongly correlated. In the case of perfect collinearity, one could not find a unique solution to the regression problem because the matrix involved is not invertible. Non-exact, but still significant, correlations between independent variables   could still produce inaccurate results. See   \cite{feigelson2012modern}, page 198, for a discussion on this matter. Although a variety of procedures have been suggested to deal with such situations, we do  not consider them here because the issue is beyond the scope  of our work. Thus, $M-V-R$ relation would not be fitted in our data analysis. 
 The data analysis presented here examines the scaling rules derived based on MOD; thus, this analysis is model dependent.

\subsection{Results}
\begin{figure}[H]
\begin{center}
\includegraphics[width=16cm]{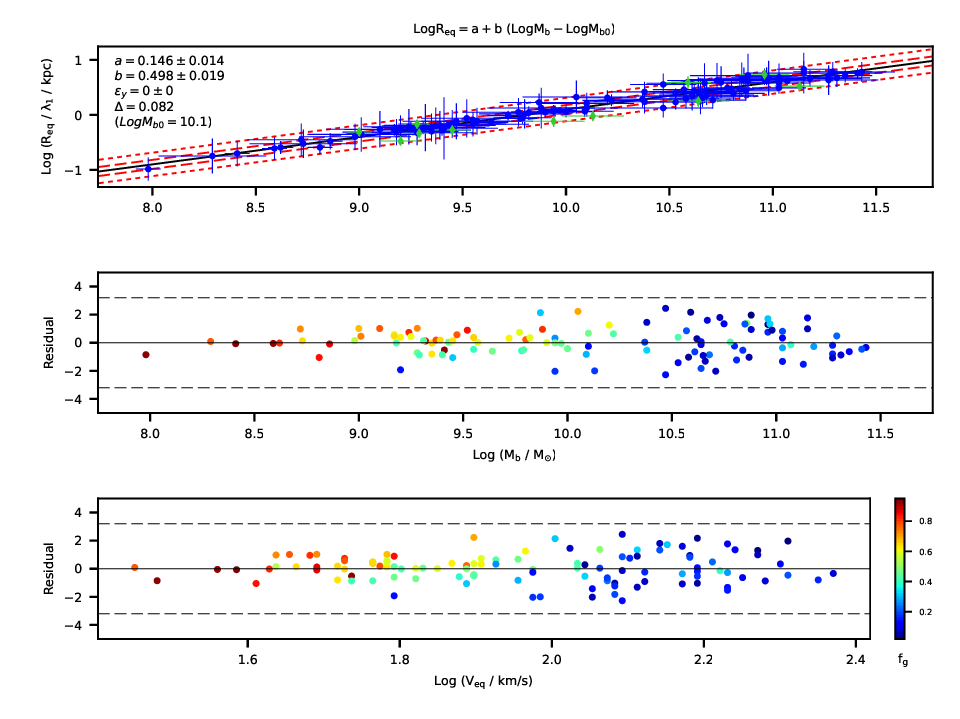}
  \vspace{-25pt}
\caption{ Data analysis of the scaling rule $R_{eq} \propto M^{0.5}$  based on the data of SPARC \citep{2016ApJ...816L..14L,2016AJ....152..157L}. In this graph it is assumed that $R_{s}=R_{eff}$ and $(~ \lambda_{2}~,\lambda_{3}~,\lambda_{4} ~)/ ~ \lambda_{1} = (~ 1.1~,0.039~,0.19 ~)$ while $\sigma_{eq}=15~km/s$. The red dashed ( red  dotted ) lines in the upper plot show the $1 \sigma $ ( $2.6 \sigma $) bands which encloses $60 \%$ ( $99 \%$ ) of the values assuming a Gaussian distribution \citep{2013MNRAS.432.1709C}. The outlier points, which are automatically excluded from the line fit,  are displayed as green diamonds.  Studentized residuals are provided in two bottom plots with Grubbs's critical line ( dashed lines )  as an indicator of outliers.  Compare this figure with Figs. \ref{fig:R2MReff}, \ref{fig:R2MRd} and \ref{fig:R2MRHI} which assume $R_{eff}$, $R_{d}$ and $R_{HI}$ as the scale length of systems ( instead of $R_{eq}$ ) respectively. The objects are color coded by gas fraction $f_{g}=M_{g}/M_{b}$.  Color online. \label{fig:R2M}}
\end{center}
\end{figure}

\begin{figure}
\begin{center}
\includegraphics[width=16cm]{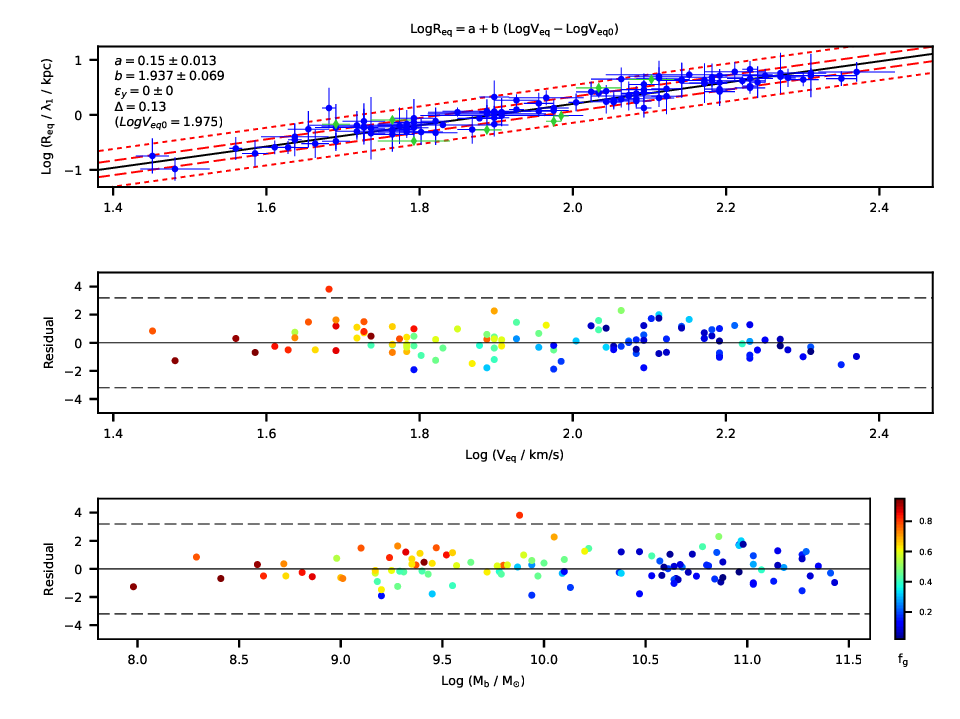}
  \vspace{-25pt}
\caption{ Data analysis of the scaling rule $R_{eq} \propto V^{2}_{eq}$. See Fig. \ref{fig:R2M} for the related parameters and explanations. Compare with Fig. \ref{fig:V2Rsig0} in which we have assumed $\sigma_{eq}=0$. Also, compare  this figure with Figs. \ref{fig:V2RReff}, \ref{fig:V2RRd} and \ref{fig:V2RRHI} which assume $R_{eff}$, $R_{d}$ and $R_{HI}$ as the scale length of systems ( instead of $R_{eq}$ ) respectively. Color online. \label{fig:V2R}}
\end{center}
\end{figure}

\begin{figure}[H]
\begin{center}
\includegraphics[width=17cm]{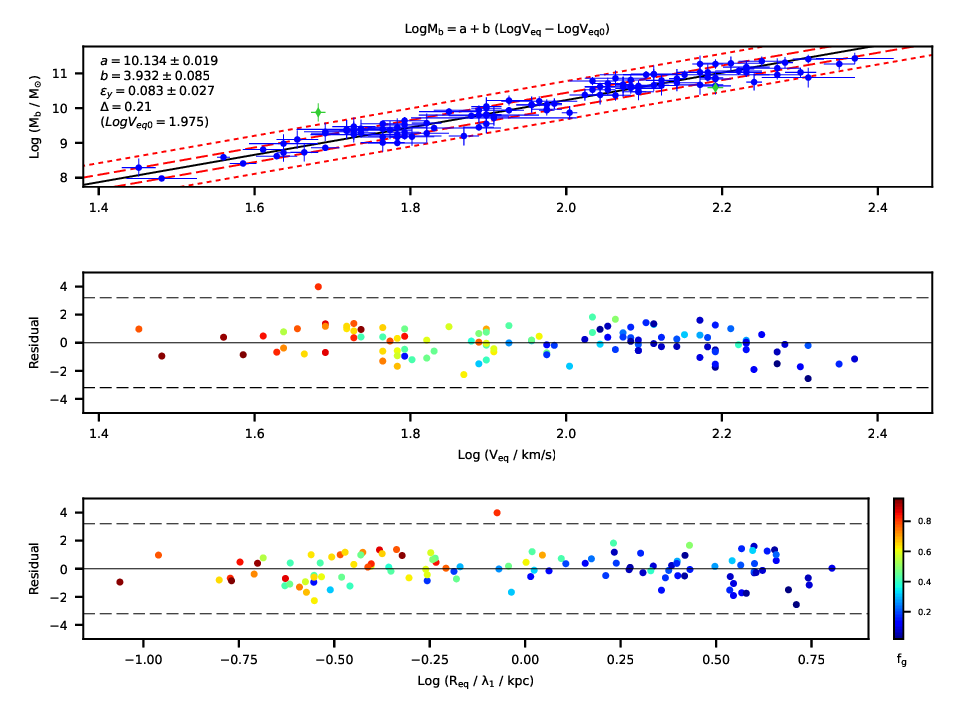} 
  \vspace{-25pt}
\caption{ Data analysis of the scaling rule $M \propto V^{4}_{eq}$. See Fig. \ref{fig:R2M} for the related parameters and explanations. Compare with Fig. \ref{fig:TFsig0} in which we have assumed $\sigma_{eq}=0$. Color online. \label{fig:TF}}
\end{center}
\end{figure}

The results of the data analysis are provided in Tables \ref{table:Reff} to \ref{table:sizes}, Figs. \ref{fig:R2M} to \ref{fig:TF} and also Figs. \ref{fig:V2Rsig0} to \ref{fig:V2RRd}. Because we have five free parameters in $R_{eq}$ and $V_{eq}$, i.e. $\lambda_{1}$ to $\lambda_{4}$ and $\sigma_{eq}$, it is quite hard to probe all of the possible values for these parameters. However, the order of the  magnitude  of $\lambda_{i}$ could be speculated by using the values of $c_{1}$, $\alpha$,   $\beta$,  $\gamma_{1}$ and $\gamma_{2}$. Thus, one could restrict the domain of possible values of the parameters. Moreover, we can temporarily decrease the degrees of freedom by dividing $R_{eq}$ by any of  $\lambda_{i}$ parameters since we only work with logarithmic values.  We choose the parameter $R_{eq}/\lambda_{1}$  which is equivalent to putting $\lambda_{1}=1$. We determine the value of $\lambda_{1}$ at the end of the data analysis by using the zero-points of the scaling rules.

\begin{table}
\begin{center}

\begin{small}   
  \begin{tabular}{|c|c|c|c|c|c|c|c|c|c|c|c|c|c|c|c|c|}   
    \hline 
     \multicolumn{4}{|c|}{ Fitting Parameters}& \multicolumn{4}{c|}{$\log R_{eq}=a+b(\log M_{b} - \log M_{b0})$} &\multicolumn{4}{c|}{$\log R_{eq}=a+b(\log V_{eq} - \log V_{eq0})$} & \multicolumn{4}{c|}{$\log M_{b}= a+b(\log V_{eq} - \log V_{eq0})$}   &  \\  \hline
    $\frac{\lambda_{2}}{\lambda_{1}}$ & $\frac{\lambda_{3}}{\lambda_{1}}$  & $\frac{\lambda_{4}}{\lambda_{1}}$ & $\frac{\sigma_{eq}}{(km/s)} $ & $a$ & $b$& $\Delta$ & $\epsilon_{y}$  & $a$ & $b$& $\Delta$ & $\epsilon_{y}$   & $a$ & $b$& $\Delta$ & $\epsilon_{y}$  &   \\
    \hline
0.80 &	0.055 &	0.24 &	17 & 0.244 & 0.496 & 0.075 & 0.0 & 0.237 & 1.917 & 0.13 & 0.0 & 10.130 & 3.981 & 0.21 & 0.085   &  \\
0.85 & 	0.056 &	0.21 &	18 & 0.215 & 0.498 & 0.076 & 0.0 & 0.211 & 1.970 & 0.12 & 0.0 & 10.128 & 4.007 & 0.21 & 0.085  & \\
1.00 & 	0.055 &	0.21 &	13 & 0.215 & 0.499 & 0.076 & 0.0 & 0.215 & 1.925 & 0.12 & 0.0 & 10.137 &  3.889 & 0.21 & 0.082  &  \\
1.10 &	0.057 &	0.22 &	16 & 0.240 & 0.491 & 0.073 & 0.0 & 0.225 & 1.933 & 0.12 & 0.0 & 10.132 & 3.956 & 0.21 & 0.084  &  \\
1.15 & 0.058 &	0.21 &	17 & 0.221 & 0.495 & 0.076 & 0.0 & 0.217 & 1.951 & 0.12 & 0.0 & 10.130 &  3.981 & 0.21 & 0.085   &  \\ \hline
  \end{tabular}
  \vspace{-12pt}
  \caption{ Results from fitting the bivariate scaling relations of Fig. \ref{fig:scaling} when the size of the stellar disk is assumed as $R_{s} = R_{d}$.  Data from SPARC \citep{2016ApJ...816L..14L,2016AJ....152..157L}.} 
\label{table:Rd} 
\end{small}
\end{center}
\end{table}

\begin{table}
\begin{center}

\begin{small}   
  \begin{tabular}{|c|c|c|c|c|c|c|c|c|c|c|}   
    \hline 
     \multicolumn{2}{|c|}{}& \multicolumn{4}{c|}{$\log R_{eq}=a+b(\log M_{b} - \log M_{b0})$} &\multicolumn{4}{c|}{$\log R_{eq}=a+b(\log V_{f} - \log V_{f0})$}    \\  \hline
    Scale   & $\frac{\sigma_{eq}}{(km/s)} $ & $a$ & $b$& $\Delta$ &  $\epsilon_{y}$ & $a$ & $b$& $\Delta$ &  $\epsilon_{y}$   \\
    \hline   
 $R_{eff}$ & 15 & 0.526 & 0.267 & 0.18 & 0.102 & 0.539 & 1.014 & 0.21 & 0.134   \\
  $R_{d}$   & 15 & 0.394 & 0.335 &  0.15 & 0.0 &  0.426 & 1.067 & 0.16 & 0.0    \\
 $R_{HI}$  & 15 & 1.214 & 0.351 & 0.11 & 0.0 & 1.216 & 1.554 & 0.16 & 0.0   \\
    \hline   
  \end{tabular}
  \vspace{-8pt}
  \caption{ Fitting the scaling relations for different scale length   $R_{eff}$, $R_{d}$, $R_{HI}$   while the velocity dispersion in not negligible ($\sigma_{eq}=15~km/s$). Related graphs are shown in Figs. \ref{fig:R2MReff} to \ref{fig:V2RRHI}. Data from SPARC \citep{2016ApJ...816L..14L,2016AJ....152..157L}. } 
\label{table:sizes} 
\end{small}
\end{center}
\end{table}

Another point is that the LTS-LINEFIT code forces the fit to a point where $\chi^{2}/(N-2) =1$ as mentioned before; thus, one could not find a single set of best parameters  based on minimizing   $\chi^{2} $. Instead, we judge the outcomes of the code  based on their residual behavior.  Many residual plots show linear trend or curvature; these results will be neglected.  We  systematically search for any pattern ( linear or second order ) in residuals and exclude them.  Tables \ref{table:Reff} and \ref{table:Rd} report the acceptable fitting parameters while the best outcome is distinguished by a $\surd$ sign in Table \ref{table:Reff}. Table \ref{table:Reff} presumes $R_{s} = R_{eff}$ in deriving the equilibrium radius $R_{eq}$ while Table \ref{table:Rd} assumes $R_{s} = R_{d}$.

The plots corresponding to the best outcome of the fit, i.e. $(\lambda_{2}, ~\lambda_{3}, ~\lambda_{4})/\lambda_{1} = (1.1,~0.039,~0.19)$ and $\sigma_{eq} = 15 km/s$ while $R_{s}=R_{eff}$,  are provided in Figs. \ref{fig:R2M} to \ref{fig:TF}. For each plot we will report Pearson's and Spearman's  coefficients as measures of linear relationship and   monotonicity of the relationship between two variables respectively \citep{press2007numerical}. Despite relatively large uncertainties in the data, specially the high error in characteristic radius, one could see that the results are close to the theoretical predictions. 
 Because of the logarithmic scales, the horizontal and vertical axes seem compressed; however, this help us to better present the results of the fit.

Fig. \ref{fig:R2M} represents the mass-size scaling rule $R_{eq} \propto M^{0.5}_{m}$. The theoretical prediction of the  slope lies within the derived bound, i.e. $b=0.498 \pm 0.019$. In addition, Pearson and Spearman correlations could be found  as 0.98 and 0.97 respectively. Moreover, in this case the intrinsic scatter is  negligible while $\Delta = 0.082 $ dex ($21~\%$). Remembering that the baryonic mass is in units of $M_{\odot}$, we could see from the intercept of this relation that
\begin{eqnarray} \label{bound1}
\frac{1}{2} \log \left( \frac{G\alpha \lambda^{2}_{1} /(2c_{1}a_{0}\beta )}{kpc^{2}/M_{\odot}}  \right) = - 4.88  \pm 0.21
\end{eqnarray}
which will be used below to estimate the parameters $\alpha$, $\beta$, $\lambda_{1}$ and $c_{1}$.

The next scaling rule, i.e. $R_{eq} \propto V^{2}_{f}$, is plotted in Fig. \ref{fig:V2R}. The slope of the plot is $b= 1.937 \pm 0.069$ while the theoretical prediction is $b=2$. Thus,  the derived value contains the theoretical prediction. Also, we mention that in this case the intrinsic scatter is  negligible and $\Delta = 0.13 $ dex. Furthermore, Pearson's and Spearman's correlations are derived to be  0.94 and 0.93 respectively. The intercept of the relation is also interesting. Regarding the units which are used in plot \ref{fig:V2R}, i.e. $V_{eq}$ in units of $km/s$ and $R_{eq}$ in units of $kpc$, we could see that an estimation for the fundamental value  of $ \beta c_{1}a_{0}$ could be derived as 
\begin{eqnarray} \label{bound2}
\log \left(\frac{4\beta c_{1}a_{0}  / \lambda_{1}}{(km/s)^{2}/kpc}\right)=3.68 \pm 0.15 .
\end{eqnarray}

The last scaling rule is the   baryonic Tully-Fisher relation which is shown in Fig. \ref{fig:TF}. The slope is derived as $b=3.932 \pm 0.085$ and it includes the theoretical slope of $b=4$. In addition, we have $\Delta = 0.21 $ dex in this case. Furthermore, BTFR shows Pearson and Spearman correlations as 0.96 and 0.95 respectively. The intercept of the plot is also found as 
\begin{eqnarray} \label{bound3}
- \log \left( \frac{8c_{1} \alpha \beta a_{0} G }{(km/s)^{4}/M_{\odot}}  \right) = -2.37  \pm 0.19
\end{eqnarray}
which lacks the presence of $\lambda_{1}$ naturally because this scaling rule is not dependent to the size.

 Conformal and Grumiller gravities, which both display linear potentials similar to MOD, are also shown to be consistent with Tully-Fisher relation \citep{2018PhLB..782..433O,2021GrCo...27..157G}. 
The Tully-Fisher relation, and especially its baryonic version,   has been much discussed in the literature.   The work by   \cite{2016ApJ...816L..14L}, in particular, is important because of their employed sample and  method of data analysis. The same code and ( almost ) the same sample is applied here. However,    \cite{2016ApJ...816L..14L} fit the BTFR by weighting each galaxy as $f^{2}_{g}$. By using this assumption, they derive a slope as $b=3.95 \pm 0.34$ and a zero point equal to $1.86 \pm 0.60$. These authors have also reported the error-weighted fit, i.e. the same method that is used here, by which they have derived $b=3.71 \pm 0.08$ and the intercept of $2.27 \pm 0.18$. See Fig. \ref{fig:TFsig0} for a similar result.  Although,  the derived slope of BTFR in Fig. \ref{fig:TF} is larger because  we have used  equilibrium velocity $V_{eq}$ in that graph instead of $V_{f}$. As a result, galaxies with smallest velocities move toward right on the velocity axis causing an increase in the slope of the BTFR line  ( as discussed in the previous section ).

 \cite{2016ApJ...816L..14L} have argued that the small intrinsic scatter in BTFR  is below the expectations from the standard model. Moreover, the residuals of the BTFR plot display no connections with galactic surface brightness or radius which is puzzling ( in the context of $\Lambda CDM$ cosmology ) regarding some semi-analytic scenarios of galaxy formation. See Fig. 2 of that paper.  

From theoretical point of view, the most prominent source of error in our analysis is the fact that we have not included the  share of galactic interactions in virial energy. In Newtonian gravity,  the interaction with other objects declines with distance; thus, it could be safely neglected in non-merging systems. However, the present model  possesses a constant acceleration $2c_{1}a_{0}$ which is present even at very large radii. This means that the effect of the neighbouring galaxies should not be neglected in general. In  \cite{Shenavar:2016xcp}, a scheme has been devised to estimate the effect of a homogeneous spherical environment on a galaxy within the sphere. The result is that if the radius of the sphere is approximately $P$ while the galaxy is at the distance $r$ from the center of the sphere, then an added acceleration of the approximate magnitude of $2c_{1}a_{0}r/P$ should be included in the analysis. However, the galaxy-galaxy interactions and especially the tidal forces have not been studied yet for MOD. By including these interactions, or at least by  estimating the error due to them, one could derive more accurate galactic  scaling rules.

\subsection{Estimating the dimensionless constants $\alpha$, $\beta$, $c_{1}$ and $\lambda_{1}$:}
 It should be mentioned that, as it is now manifest from Eqs. \ref{bound1} to \ref{bound3}, the four constants $\alpha$, $\beta$, $c_{1}$ and $\lambda_{1}$ participate only in zero points of the scaling rules. Strictly speaking, they have no share in determining the overall behavior  of the rules, i.e.  the exponent of $V$, $R$ and $M$ in the scaling rules. In fact, the same is true for $G$ and $a_{0}$. 

Because the uncertainties in constraint equations \ref{bound1} to \ref{bound3} are quite substantial, one could not pin point the values of  $\alpha$, $\beta$, $c_{1}$ and $\lambda_{1}$. Thus, we confine the possible values of these parameters as follows: The parameter  $c_{1}$ is supposed to be within $0.01< c_{1}< 0.09$ which is suggested by the prior works \citep{Shenavar:2016bnk,Shenavar:2016xcp,shenavar2018local}. Moreover, $\alpha$ and $\beta$ are allowed to have a $3 \%$ tolerance about the values reported in Tab. \ref{table:paras} for the thick disk. Also, we presume $0.7<\lambda_{1}<2.0$ which is consistent with our estimation for this parameter  based on $\lambda_{1}=\alpha/ (4\beta^{\prime} c_{1})$.  Assuming these, one could solve Eqs. \ref{bound1} to \ref{bound3}  using MATHEMATICA to obtain $\lambda_{1}=0.74$, $\alpha = 0.27$, $1.43< \beta < 1.52$ and $ c_{1} = 0.035/ \beta $. This completes our data analysis. 

\section{Discussions on the Results}
\label{Discussion}
\begin{enumerate}

\item 
 It is sometimes argued that BTFR is a "fundamental" relation, i.e.  a new law of nature probably resulting from a MOND-like paradigm, while mass-size and velocity-size relations are considered as the end result of galaxy formation.  In other words, according to this viewpoint,  $M-R$  and $V-R$ relations are considered as "normal" galaxy scaling relations which arise due to collective behavior  of particles in  systems  while BTFR points to a fundamental law of physics. This argument is mostly based on the observed small scatter of BTFR  and its lack of residual correlations which are reported by   \cite{2016ApJ...816L..14L,2019MNRAS.484.3267L}. See also   \cite{2017MNRAS.472L..35D} for discussions on the scatter in  BTFR and its curvature.  However, the presumption that BTFR and  $M-R$ ( or $V-R$ ) are fundamentally different  is not necessarily true because, as we have shown in Sec. \ref{GSRV} and   \ref{Derive} , the same set of scaling rules could be derived from a modified Poisson equation as well as virial theorem. Thus, they could be achieved by assuming a new fundamental law, i.e. a modified Poisson equation, or as a result of combined effects of all particles ( through virial theorem, for example ). In fact, it seems more plausible to presume that if  a new law of gravity is assumed, then the effects of this new law would be observable in all galactic phenomena ( including galaxy formation ) not just one of them. Interestingly, Milgrom derives BTFR from the equation of motion \citep{Milgrom:1983pn} and also as a virial relation based on the  scale invariance of the theory  \citep{2009ApJ...698.1630M}; however, the  measure of velocity in this latter approach is the mean squared velocity.
 
 Moreover, the results of Tables \ref{table:Reff} and \ref{table:Rd} shows that the intrinsic scatter $\epsilon_{y}$ in  both $M-R$  and $V-R$ relations is smaller than BTFR. The same is true about $\Delta$. However, Table \ref{table:sizes} shows that if we assume $R_{eff}$ as the scale length of the whole system,  then  $\epsilon_{y}$ in  $M-R$  and $V-R$ relations is larger than BTFR. \textit{ This observation probably suggest that $R_{eff}$  is not a good size measure of total baryonic mass ( see \ref{conc1} below ) rather than the fundamentality of BTFR in comparison to $M-R$  and $V-R$.}

On the other hand, in the context of dark matter models, the Tully-Fisher relation has mostly been considered  as a result of galaxy formation.   \cite{1998MNRAS.295..319M}, for example, propose a scenario to build Tully-Fisher relation based on an underlying correlation between virial properties of halo, namely  the virial mass $M_{200}$ and virial velocity $V_{200}$ of the DM halo. A $V-R$ relation is also suggested by these authors.

\item 
   As the present results show, the exponents  in $M-R$  and $V-R$  relations critically depend   on the definition of the galactic size $R$. For example, the exponent in $R \propto M^{n}$ relation could be derived as $n = 0.498$ for $R=R_{eq}$ while we have $n= 0.335, ~ 0.267,~ 0.351$ for $R= R_{d}$, $ R_{eff}$  and $ R_{HI}$ respectively. See Tab. \ref{table:sizes}.  See also   \cite{2005PhRvL..95q1302M} who derives $R_{d} \propto M^{0.29 \pm 0.03}_{b}$ and also $R_{P} \propto M^{ 0.16 \pm 0.04}_{b}$. Here, $R_{P}$ is the radius where the peak of the baryonic rotation velocity occurs.  Thus, the issue of suggesting a proper scale for the systems could not be under-estimated.

Interestingly, a recent paper by \cite{2022MNRAS.512.2697R} derives the slope and the intercept of the observed HI size-mass relation of 204 galaxies respectively  as $ 0.501 \pm 0.008$ and$-3.252^{+0.073}_{-0.074}$ ( the observed scatter of $0.057~ dex $ ). Their sample is derived  from the MIGHTEE Survey Early Science data  which benefits from the high sensitivity of MeerKAT and contains galaxies within one billion years in lookback time ( $ z \leq 0.084$ ). The data set and the method of fitting in this work is different from ours; however, the slope of this tight relation is pretty close to the value derived here for the baryonic size-mass relation. Since the horizontal and vertical axes in the HI size-mass relation are not similar to those of the baryonic size-mass relation,  the intercepts of the two relations are distinct.

The scaling rules of    \cite{2017ApJ...836..151S} are derived based on  well-motivated observational evidence ( discussed above ). Thus,  any significant deviation  in the exponent of $M-R$  or $V-R$ relations from those of Schulz's could be interpreted  in one of the following ways: First, it could be concluded that Schulz's assumptions needs to be revised ( to include the share of dark matter or the effects of a modification in gravity ). Second, one could infer that the scale length used in data analysis does not properly describe the whole system. The present work shows that Schulz's scaling rules could be achieved within MOD while the virial equilibrium scale length  $R_{eq}$ fits well enough into the scaling relations.

\item \label{conc1}
 In BTFR, $M_{b}$ represents the total ( gas + stellar ) baryonic mass. Our analysis shows that,  in the same way, any scale length of the system should represent  the scale of both gaseous and stellar  components. In other words, a scale length representing only stellar distribution ($R_{d}$ or $R_{eff}$) could not be a suitable length especially in a sample like SPARC in which many systems show substantial gas fractions ( $f_{g} $ is up to 0.95 $\%$ in SPARC ).

\item 
 Here we introduced a scale length which includes the stellar and gas lengths with their proper share which depends on $f_{g}$. The relevance of another scale length, namely the truncation radius $R_{t}$, to the scaling rules needs also to be examined because this scale length too depends on stellar and gas distributions. Although, we do not pursue this case here, we only mention that the truncation radius could be derived by extending the method of   \cite{1983MNRAS.203..735C} to the modified potential of \ref{secondorder}. We will report the truncation radii for a sample of objects  ( as a part of fitting rotation curves in MOD ) in future. The reliability of $R_{t}$ in fitting scaling rules will be checked too.

\item
 The inclusion of the velocity dispersion $\sigma_{eq}$  changes the exponent in $M-V$ and $R-V$ relations considerably. By presuming a negligible velocity dispersion, the exponents in these two relations deviate manifestly from theoretical predictions. To see this, compare the results of Figs. \ref{fig:V2R} and \ref{fig:TF} which presume $\sigma_{eq}=15~km/s$ with Figs. \ref{fig:V2Rsig0} and \ref{fig:TFsig0} for which $\sigma_{eq}=0$. This point probably refers to the statistical nature of the scaling relations because the relevance of $\sigma_{eq}$ could be simply justified from the virial theorem.

\item 
  \cite{2019MNRAS.484.3267L} consider the Fall relation \citep{2012ApJS..203...17R} which is a relation between specific angular momentum versus mass. This is essentially an $M-V-R$ relation. See   \cite{2012ApJS..203...17R} and references therein.  Lelli et al. argue that the Fall relation is  about five times broader than the stellar-mass TF relation. Thus, they conclude that the TF relation must be more fundamental compared to  Fall, because TF is tighter and shows no residual dependency  to additional variables, i.e. scale length. The broadening in Fall relation is argued to originate from the vertical scatter in the mass-size relation ( e.g. see Fig. 2 of \cite{2016AJ....152..157L} which relates $R_{eff}$ and $R_{d}$ to  $[3.6]$ luminosities $L_{[3.6]}$ ). 

However, Lelli et al.   continue  to  argue that : "the BTFR is already the Fundamental Plane of galaxy disks: no value is added with a radial variable as a third parameter". Our results encourages a disagreement on this conclusion and our line of reasoning is as follow: The result of plane-fitting, tangled with the issue of collinearity in the present case,  is  intrinsically more complicated than line fitting  and possibly even imprecise ( depending on the degree of collinearity ). Therefore, the outcome of such a plane fit is  hard to interpret. However,  the weakness of a three-parameter relation ( such as Fall's ) does not rule out the existence of other bivariate correlations,   such as $R_{eq} \propto V_{eq}^{2}$ and $R_{eq} \propto M_{b}^{0.5}$, as the  present work shows.  

Thus, based on the tightness of $R-V$ and $M-R$ relations reported here, we conclude that BTFR is not necessarily equal to the fundamental plane. In fact, our results supports the other conclusion of  \cite{2019MNRAS.484.3267L} who suggest "considering simultaneously the BTFR and the mass-size relation of galaxies provides a stronger constraint on galaxy formation models than considering the Fall relation alone". This is quite a good summary of the situation. Indeed, it could be further argued that the set of  $ \left\lbrace M-V,~M-R \right\rbrace$, as the governing scaling rules of the system, is even superior to the set of $ \left\lbrace M-V, ~M-V-R \right\rbrace$  for  the simple reason of the  collinearity issue of the latter case.

\item
The effects of a negligible velocity dispersion and also changing the scale length of systems are shown in Figs. \ref{fig:V2Rsig0} to \ref{fig:V2RRHI}. In all cases, we observe  significant lower slopes compared to the predictions of Schulz's proposal. Although Schulz's scaling rules could be derived from a modified Poisson equation too ( as proved in   \ref{Derive} ),  it is possible  that the  efficiency of $R_{eq}$ and $V_{eq}$ - which are obtained based on statistical assumptions -   in fitting scaling relations hints toward the statistical nature of these relations. This is particularly evident from BTFR because without adding $\sigma_{eq}$, the slope remains about 3.747. See Fig. \ref{fig:TFsig0} below. The deviation of this slope from the theoretical prediction of 4 is almost four times compared to when we assume $\sigma_{eq}=15~km/s$.  

\item
In the context of the dark matter paradigm, there have been reports of close connection between the properties of the luminous and the dark matter. Among the more recent works, we should  mention studies by  \cite{10.1093/mnras/stw3055} on dwarf disk galaxies and   \cite{10.1093/mnras/stz2700} on LSBs. The latter paper, for example, compares the compactness of the dark vs. luminous matter distributions and comes to conclusion that there is a strong correlation between the two. This effect could be seen as another aspect of the so-called disk-halo conspiracy. See  \cite{2010dmp..book.....S} for a review and more details. Within the dark matter point of view, such strong connection might be interpreted as \textit{ "the existence of non-standard interactions between the luminous matter and the dark matter, or non-trivial self-interaction in the DM sector or a (hugely) fine-tuned baryonic feedback on the collisionless DM distribution"} \citep{10.1093/mnras/stz2700}. Within the context of modified gravity/dynamics, however, one could discard the disk-halo conspiracy by introducing a new fifth force to explain the galactic systematic.

\item
\textbf{Cosmic evolution of the galactic scaling relations:} The evolution of the scaling relations with redshift could be checked by employing large galactic data at different redshifts. To do so, one needs to investigate the relations at fixed $M$, $R_{eq}$ or $V_{eq}$. For example, at fixed mass, one could easily see from the scaling relations of Fig. \ref{fig:scaling} that
\begin{subequations}
\begin{align} \label{RfixedM}
 R_{eq}  &\propto  \left(~ c_{1}(z) H(z)~ \right)^{-0.5} && \text{Fixed}~M \\ \label{VfixedM}
 V_{eq} &\propto  \left(~  c_{1}(z)H(z) ~ \right)^{0.25}  && \text{Fixed}~M 
\end{align} 
\end{subequations}
in which we have re-entered the Neumann parameter $ c_{1}(z)$ as an evolving function of redshift $z$. The  equation  \ref{RfixedM} predicts that the size of galactic systems should decrease with increasing redshift. This relation has not been checked yet, since the equilibrium size $R_{eq}$ is introduced here for the first time. However, various observational reports based on different definitions of size agree qualitatively ( though not always quantitatively ) that  in fixed mass, the  galactic size reduces with redshift \citep{2012ApJ...756L..12M,2014ApJ...788...28V,2019ApJ...880...57M}. Also, the  equation  \ref{VfixedM} anticipates that, at fixed $M$, the value of $V_{eq}$ should increase  as a function of $z$. The redshift evolution of the baryonic Tully-Fisher relation, in particular, is expected to be investigated in upcoming surveys \citep{2012IAUS..284..496H} while a positive evolution of the BTFR zero point from $z \sim 0.9$ to $z \sim 2.3$ has already been reported \citep{2017ApJ...842..121U}.

See also  \cite{2022MNRAS.512.2697R} who derive  the HI size-mass relation for two
redshift bins of $ z \leq 0.04$ and $ 0.04 < z \leq 0.084$. For these two bins, they report the slopes to be $0.485^{+0.012}_{-0.011}$ and $0.526^{+0.013}_{-0.013}$  respectively ( See Table. 2 of that paper ). Within this considerably small redshift range, the authors conclude marginal difference between the slope
and intercept of the two subsamples.

Moreover, at fixed equilibrium radius $R_{eq}$, one could expect
\begin{subequations}
\begin{align}
 M &\propto  ~c_{1}(z)H(z)~  && \text{Fixed} ~ R_{eq} \\
 V_{eq} &\propto  \left(~ c_{1}(z)H(z) ~ \right)^{0.5} && \text{Fixed} ~ R_{eq}.
\end{align} 
\end{subequations}
In addition, by keeping the equilibrium velocity $V_{eq}$ as a constant, one could derive
\begin{subequations}
\begin{align}
 M &\propto  \left(~ c_{1}(z)H(z)~ \right)^{-1} && \text{Fixed} ~ V_{eq} \\
 R_{eq} &\propto  \left(~ c_{1}(z)H(z)~ \right)^{-1} && \text{Fixed} ~ V_{eq}.
\end{align} 
\end{subequations}

It is worth noting that  these six relations could be in fact used  to check the consistency of the whole model since they all predict a certain behavior for the parameter $c_{1}(z)H(z)$. In other words, one expects that the following relations
\begin{eqnarray}
c_{1}(z)H(z) &\propto R^{-2}_{eq}(z)|_{M=cte}  &\propto V^{4}_{eq}|_{M=cte}        ~~~\propto (1+z)^{\eta} \\
c_{1}(z)H(z) &\propto M(z)|_{R_{eq}=cte} &\propto V^{2}_{eq}|_{R_{eq}=cte}   ~~~\propto (1+z)^{\eta} \\
c_{1}(z)H(z) &\propto M^{-1}(z)|_{V_{eq}=cte}  &\propto R^{-1}_{eq}|_{V_{eq}=cte}  ~~~\propto (1+z)^{\eta}
\end{eqnarray}
predict a more or less constant $\eta$. A sharp inconsistency in the values derived for $\eta$ could either rule out the model, the stability condition proposed here ( especially the steady state condition $\ddot{I}=0$ presumed in Sec. \ref{GSRV} )  or both.

On the other hand, if the parameter $\eta$ was derived consistently, then one could take the model more seriously and constrain the evolution of $c_{1}(z)H(z)$ which is much needed in MOD cosmology ( This information could provide useful clue on the behavior of the modified potential of MOD and, consequently, the form of the function $\zeta (a)$ in   \cite{2018arXiv181005001S}. ). The issue of the cosmic evolution of  scaling relations needs careful data analysis, based on large sample of data, and it is beyond the scope of the current work.

\end{enumerate}

\section{Conclusion}
\label{Conclusion}  
In this work, the slopes and the zero points of the  scaling rules ( summarized in Fig. \ref{fig:scaling} )  are found to be close to the theoretical expectations of MOD. However, more accurate data in future could further decrease the uncertainty in the scaling rules.  Especially by reducing the error in   distance measurements one could expect a significant improvement in data analysis. 

Comparing with scaling rules inferred based on dark matter hypothesis, which are naturally dependent to presumed halo's characters \citep{1997gsr..proc....3W,2019ApJ...877...64D}, one could see that Schulz scaling laws are exceptional in the sense that they solely relate observable quantities. Moreover,  in comparison with  the work of  \cite{2018arXiv181102025N}, which presumes  self-similar nature of NFW halo plus  highly nonlinear galactic-halo relations of mass and size derived from sophisticated large-scale simulations, one could see that the simplicity of the virial method ( presented here ) is quite noticeable.  Furthermore, the success of these rules is a possible indication toward modified dynamics. 

There are two different procedures toward the scaling relations in physics. According to the first method, one investigates the scaling laws based on pure dimensional analysis plus  a few phenomenological or theoretical facts.  The analysis which resulted in Fig. \ref{fig:accsigma},  based on the pioneer work of   \cite{2012LRR....15...10F}, is an example of this approach. As another case, one could see the  derivation of the scaling laws provided in   \ref{Derive}. However, the scaling relations could also be approached based on rigorous statistical/thermodynamical analysis of the system, i.e. as a result of the "collective behavior" of different components in the system. The analysis presented in Secs. \ref{GSRV} and \ref{Measures} is an example of this second approach toward the scaling relations. 

Also, this work is a first step toward a statistical view toward gravitational systems in a modified dynamical model; however, it is far from being  conclusive. A more elaborated approach to the statistical  mechanics of galactic systems using the modified Poisson equation \ref{secondorder} could improve the present work at least from theoretical point of view. Especially, because  the long range interaction due to the modified term in Eq. \ref{Potential} is stronger than  Newtonian one, the statistical behavior of the gravitating systems could be quite different from pure Newtonian systems.
 See \cite{2022arXiv220502281S} for a few short comments on the statistical mechanics of gravitating systems in MOD.  

The above analysis showed that by maximizing the scalar virial energy, which is equivalent to minimizing the kinetic energy, one could successfully derive the galactic scaling rules.  However, the tensor virial theorem contains more information about the system compared to the scalar one \citep{1978vtsa.book.....C}. Thus, an appropriate future investigation could start from the implications of tensor virial theorem and then compare the results with observations \citep{2007MNRAS.379..418C,2012MNRAS.422.1767A}.  It is also desirable to know whether there is any condition in three dimensions equivalent to maximization of scalar virial  energy in one dimension. If there is such  condition, it could lead to favoring some mass configurations, such as flattened systems, over others.

Moreover, a natural question now is whether we could justify a baryonic Faber-Jackson  relation \cite{1976ApJ...204..668F} using modified potential \ref{Potential}. The original work of Faber $\&$ Jackson     relates the random velocities of  elliptical galaxies with their absolute luminosities. However detailed study of this matter is beyond the scope of the present work,   \cite{2012LRR....15...10F} have argued that the zero point of Faber-Jackson  relation is also proportional to $Ga_{0}$ which could be considered as another sign for a modified dynamics at work at galactic scales.  

The scaling rules which are discussed here predict a certain redshift dependency of mass, velocity and size when one of these parameters is presumed fixed. This redshift dependency of the main parameters is probably hard to measure in an accurate way; however, any future results based on this method could constrain the evolution of $c_{1}(t)H(t)$. Eventually, this knowledge could be helpful in investigating the MOD cosmology.

Scaling relations still show surprising results from both theoretical and observational points of view. For example,   \cite{2018MNRAS.480L..23R,2020MNRAS.491.4843R,2020MNRAS.499.5656R} proposes that the specific angular momenta of stars, atomic and molecular gas are tightly related to their masses and velocity dispersions via a self-regulation process driven by disk gravitational instability. See also   \cite{2021A&A...647A..76M} for a discussion on the baryonic specific angular momentum of disc galaxies which extends the Fall relation by 2 orders of magnitudes towards lower masses. Moreover,   \cite{2021A&A...651L..15M} shows that if one incorporates the gas content to the Fall relation, one finds a much lower intrinsic scatter,  for a tight angular-momentum plane for disc galaxies. In addition, although the BTFR  is shown to be in agreement with dwarf irregular galaxies within LITTLE THINGS sample ( see   \cite{2017MNRAS.466.4159I} ),   \cite{2019ApJ...883L..33M,2020MNRAS.495.3636M} report a deviation from this scaling relation for a few isolated gas-rich ultra-diffuse galaxies. These cases   certainly need a thorough analysis within the framework of modified dynamical models.

\section*{Acknowledgements}
I would like to thank  Stacy S. McGaugh, Benoit Famaey, Federico Lelli and James M. Schombert for providing their data freely. Also, I thank Michele Cappellari and his colleagues for providing LTS-LINEFIT code which is available here \url{https://www-astro.physics.ox.ac.uk/~mxc/software/}. I am thankful to Hossein Afsharnia for his help in understanding LTS-LINEFIT code and checking some results during this work. I also appreciate helpful discussions and comments by Kurosh Javidan, Earl Schulz, Neda Ghafourian, Mahmood Roshan, Pavel Mancera Pi{\~n}a, Alessandro Romeo and  Paolo Salucci.  Mohammad Moghadassi kindly guided me with  tikz package for which I am   grateful.   Furthermore, I acknowledge the anonymous reviewer whose comments helped to clarify and strengthen the discussions and presentation of the work.  This research has made use of NASA' Astrophysics Data System.

\appendix
\section{Deriving the Scaling Rules from the Modified Poisson Equation}
\label{Derive}
In addition to the  derivation of the scaling laws presented in Sec. \ref{GSRV}, it is also possible to obtain these rules based on dimensional analysis of the modified Poisson equation \ref{secondorder}. This derivation is somehow essential to our discussions because it shows that the scaling rules could be obtained from the fundamental laws as well as the virial theorem. See Sec. \ref{Discussion} for the related discussions. 

We may start our analysis by considering a highly flattened, axisymmetric and self-gravitating system. One could use  the original method of   \cite{1963ApJ...138..385T} ( see also   \cite{2008gady.book.....B}, pages 103 - 107 ) to integrate Eq.  \ref{fourthorder} and derive 
\begin{eqnarray} \label{rc}
v^{2}_{c}(R)=R\frac{\partial \Phi}{\partial R}= R\int^{\infty}_{0} k dk J_{1}(kR) \left(  \frac{2\pi G k^{2}-\frac{8\pi c_{1}a_{0}}{M}}{ k^{2}} \right)   \int^{\infty}_{0}R^{\prime} dR^{\prime} J_{0}(kR^{\prime})\Sigma(R^{\prime}).
\end{eqnarray}
The details of the derivation of equation \ref{rc} will be presented in a future work; however, you may find a similar calculation in Shenavar $\&$ Ghafourian \citep{shenavar2018local}. Here, $v_{c}(R)$ is the circular velocity at radius $R$, $k$  is the wavenumber, $J_{i}$ is the Bessel function of the first kind and $\Sigma(R^{\prime})$ is the surface density at radius $R^{\prime}$.

The equation  \ref{rc}  contains two successive Hankel transforms of zeroth and first order. By inverting these Hankel transforms one could readily find the surface density $\Sigma(R)$ as follows:
\begin{eqnarray}   \label{sigma}
\Sigma(R)=\frac{1}{2\pi G}\int^{\infty}_{0} k dk J_{0}(kR) 
\left( \frac{k^{2}}{ k^{2}-k^{2}_{f}} \right)  \int^{\infty}_{0} dR^{\prime} v^{2}_{c}(R^{\prime}) J_{1}(kR^{\prime})
\end{eqnarray}
in which $k^{2}_{f} \equiv \frac{4 c_{1}a_{0}}{MG}$  is a boundary wavenumber which plays a significant role in the stability of both fluid and stellar systems  \citep{shenavar2018local}.

As mentioned above, Eqs. \ref{rc} and \ref{sigma} would be thoroughly derived and discussed in a separate work. For example, we will show that Eq. \ref{sigma} leads to the truncation of mass distribution while the same does not happen under Newtonian gravity. The saturation of disk in MOD occurs because the gravitational force in this model is significantly enhanced compared to the force provided by different halo profiles. Now, let us name the truncation radius as $R_{t}$. From Eq. \ref{sigma} it is evident that the truncation radius $R_{t}$ should depend on all baryonic ( stellar + gas ) mass. Thus it could be considered as ascale length of the systems. See Sec. \ref{Discussion} for a discussion. 

The goal of this part is to show that the scaling relations could be derived from Eq. \ref{sigma}  too. To display this, we could see that since we have $k^{2}_{f} = \frac{4 c_{1}a_{0}}{MG}$ for the singular point of Eq. \ref{sigma}, one could readily see  that $k_{f} \propto \sqrt{1/M}$. On the other hand, by assuming that the scale length of the system $R_{t}$ is inversely proportional to the boundary wave number $k_{f}$, one could simply find
\begin{eqnarray} \label{dimrm}
R_{t} \propto M^{1/2}.
\end{eqnarray} 
Moreover, one could see that the LHS of Eq. \ref{sigma} would be proportional to $M/R^{2}_{t}$ while the RHS is proportional to $V_{f}^{2}/R_{t}$ in which $ V_{f} $ is the flat rotation velocity. Thus, it is possible to see the next scaling rule
\begin{eqnarray} \label{dimrvm}
M \propto    R_{t} V^{2}_{f}
\end{eqnarray}
which, one could also derive in Newtonian gravity, as   \cite{1960ApJ...131..293B} had shown. Eliminating $R_{t}$ between these two last equations could provide BTFR
\begin{eqnarray}
M \propto V^{4}_{f}
\end{eqnarray}
while by eliminating $M$, one would obtain
\begin{eqnarray}
R_{t} \propto V^{2}_{f}.
\end{eqnarray}
 The  coefficients of proportionality for these four scaling rules must be achieved by using the constants of the theory, i.e. $G$ and $a_{0}$, as   \cite{2017ApJ...836..151S} has explained. 

Using the method of    \cite{1983MNRAS.203..735C}, we will derive $R_{t}$ and its uncertainty in a future work as a part of fitting rotation curve data. Then we could  check the viability of the truncation radius for the scaling rules. Therefore, while this second method reveals the form of the scaling laws, the method of Casertano provides the necessary machinery to analyze them.

\section{The  Virial Energy of a Compound System }
\label{Comound}
To derive an appropriate  scale length  corresponding to the virial equilibrium of a gas + stellar system, one needs to evaluate the virial energy of such compound configurations. The total virial energy of this two-component system is as follows 
$$W_{tot} \equiv -\int d^{3}\vec{x} ( \rho_{s} + \rho_{g} )  \vec{x}.\vec{\nabla} \Phi_{tot} $$  
in which 
$$ \Phi_{tot} = \Phi_{tot}^{N} + \Phi_{tot}^{MOD} $$ 
is the total potential,  
$$ \Phi_{tot}^{N} = -G \int \frac{( \rho_{s} + \rho_{g} )d^{3}\vec{x^{\prime}}}{| \vec{x^{\prime}} - \vec{x}|} $$ 
is the total Newtonian potential and 
$$ \Phi_{tot}^{MOD}=  \frac{2c_{1}a_{0}}{M}\int ( \rho_{s} + \rho_{g} ) d^{3}\vec{x^{\prime}}|\vec{x^{\prime}} - \vec{x}| $$ 
is the total potential due to the modification in dynamics. Also, here we have shown the stellar and gas components by $s$ and $g$ indices respectively. It could be seen now that in a two-component system, the virial energy consists of four different energies: the virial energy of stars due to stars $W_{ss}$, the virial energy of gas due to gas $W_{gg}$, the virial energy of stars due to gas $W_{sg}$ and finally the virial energy of gas due to stars $W_{gs}$. For example, the first virial energy  is equal to 
$$ W_{ss} \equiv -\int d^{3}\vec{x} \rho_{s}   \vec{x}.\vec{\nabla} ( \Phi^{N}_{s} + \Phi^{MOD}_{s} )   $$
 in which
  $$\Phi^{N}_{s} = -G \int \frac{  \rho_{s}d^{3}\vec{x^{\prime}}}{| \vec{x^{\prime}} - \vec{x}|}  $$ 
  and  
   $$\Phi^{MOD}_{s} =\frac{2c_{1}a_{0}}{M}\int \rho_{s} d^{3}\vec{x^{\prime}}|\vec{x^{\prime}} - \vec{x}| $$ 
are the Newtonian and MOD potentials due to stars respectively. In the same way,  the third virial energy is equal to 
$$ W_{sg} \equiv -\int d^{3}\vec{x} \rho_{s}   \vec{x}.\vec{\nabla} ( \Phi^{N}_{g} + \Phi^{MOD}_{g} )  $$
 where the terms
  $\Phi^{N}_{g} + \Phi^{MOD}_{g}$
   represent the Newtonian plus MOD potentials of the gas. 

  It has been proved in \cite{Shenavar:2016xcp} ( Appendix ) that  for typical single-component  mass distributions - i.e. homogeneous spheres, exponential spheres, exponential disks etc. -   the  virial energy has a unique dimensional form similar to Eq. \ref{Virenergy}.  Using this result, and showing the scale length of stellar and gas components by $R_{s}$ and $R_{g}$ respectively, the virial energy of stars due to stars (or the  virial energy of gas due to gas ) could be found as  $W_{ss}=-G\alpha M^{2}_{s}/R_{s} -2c_{1}a_{0} \beta M^{2}_{s}R_{s} /M $ ( or  $W_{gg}= - G\alpha M^{2}_{g}/R_{g} -2c_{1}a_{0} \beta M^{2}_{g}R_{g}/M$ for the virial energy of  gas due to gas). For the sake of simplicity, we have assumed that both systems have the same geometry; thus, they both have the same virial coefficients $\alpha$ and $\beta$. This is not necessarily true; however, this simple toy model helps to reduce  the complexity of the data analysis by decreasing the free parameters of the equilibrium radius. 
  
  In the same way, one could try to evaluate the virial energy of stars due to gas and vice versa. However, the calculations  prove to be very complicated for exponential mass distributions. The reason is the simultaneous appearance of  exponential and Bessel functions with different scale length ($R_{s}$ and $R_{g}$ ). These virial integrals do not result in simple forms based on powers of $R_{s}$ and $R_{g}$.

\begin{figure}
\begin{center}
\begin{tikzpicture}
\shade[ball color = gray!40, opacity = 0.2] (0,0) circle (3cm);
\draw[very thick,->,>=stealth] (0,0,0) -- (4,0,0) node[anchor=north east]{$x$};
\draw[very thick,->,>=stealth] (0,0,0) -- (0,4,0) node[anchor=north west]{$y$};
\draw[very thick,->,>=stealth] (0,0,0) -- (0,0,7) node[anchor=south]{$z$};
\draw[thick,->,>=stealth] (0,0,0) -- (0.5,1.3,0) node[anchor=south]{$P_{2}$};
\draw[thick,->,>=stealth] (0,0,0) -- (-2.0,1.8,0) node[anchor=south]{$P_{1}$};
\draw (0,0,0)-- (1,-1.46,0);
\node [text=blue] at (0.7,-0.6,0) {\footnotesize $R_{s}$};
\draw (0,0,0)-- (-2.83,-1,0,0);
\node [text=blue] at (-2.2,-0.55,0) {\footnotesize $R_{g}$};
\clip (0,0) circle (1.8cm);
\pgfmathsetseed{24122015}
\foreach \p in {1,...,300}
{ \fill (2*rand,2*rand) circle (0.025);
}
\end{tikzpicture}
\end{center}
  \vspace{-20pt}
\caption{A homogeneous spherical stellar distribution with radius $R_{s}$ shown by random dots is encompassed within a homogeneous spherical gas distribution which is shown as a gray sphere with radius $R_{g}$.}
  \label{fig:stargas}
\end{figure}
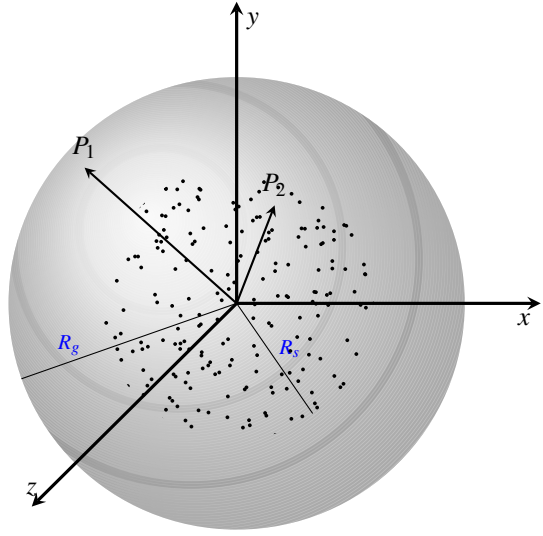

Although, it is still possible to evaluate $W_{gs}$ and $W_{sg}$ for homogeneous spherical stellar and gas distributions. Then, we will show that the terms due to the particular mass distribution of the system is negligible. In other words, the leading term in both $W_{gs}$ and $W_{sg}$ is independent of the especific geometry. Thus, we derive a general approximate form for  these two terms up to some constants. To see this, consider a  system which consists of stars and gas while both components are homogeneous and spherical. See Fig. \ref{fig:stargas}. In such system, the Newtonian virial energy of gas due to stars is as follows:
$$ W^{N}_{gs} \equiv \int \rho_{g}d^{3}x~ \vec{x}.\vec{\nabla}\Phi^{N}_{s} $$
Simple investigation of our sub-sample of SPARC shows that for all objects  we have $R_{s}<R_{g}$ as depicted in Fig. \ref{fig:stargas}.  Thus, the Newtonian potential due to stars $\Phi_{s}^{N}$, or more practically the Newtonian acceleration $\nabla \Phi_{s}^{N}$, should be evaluated for $R<R_{s}$ and $R>R_{s}$ separately
$$ W^{N}_{gs} = \int^{R_{s}}_{0} \rho_{g}d^{3}x~ \vec{x}.\vec{\nabla}\Phi^{N}_{s, in} + \int^{R_{g}}_{R_{s}} \rho_{g}d^{3}x~ \vec{x}.\vec{\nabla}\Phi^{N}_{s, out} $$
in which $\Phi^{N}_{s, in}$ ($\Phi^{N}_{s, out}$) represents the Newtonian potential due to stars inside (outside) the stellar population.  Using Gauss's law we have
 $$ W^{N}_{gs} = -4 \pi GM_{s}\rho_{g} \left(  \int^{R_{s}}_{0}\frac{R^{4}}{R^{3}_{s}}dR    + \int^{R_{g}}_{R_{s}} R dR  \right)$$
Which could be simply solved to obtain
 $$ W^{N}_{gs} =   -\frac{3GM_{s}M_{g}}{2R_{g}} \left( 1-\frac{3R^{2}_{s}}{5R^{2}_{g}}  \right).$$

In the same way, the Newtonian  virial energy of stars due to gas
 $$ W^{N}_{sg} \equiv \int \rho_{s}d^{3}x~ \vec{x}.\vec{\nabla}\Phi^{N}_{g} $$
could be found as 
 $$ W^{N}_{sg} = \int^{R_{s}}_{0} \rho_{s}d^{3}x~ \vec{x}.\vec{\nabla}\Phi^{N}_{g, in}  $$
which consists of only one integral. The reason is that   for the sample under consideration, all stellar components are encompassed within the gas; thus, the gaseous component produces no virial energy on stars outside the gas because there is no star outside $R_{g}$. Again, by deriving  $ \vec{\nabla}\Phi^{N}_{g, in}$ from Gauss's law
 $$ W^{N}_{sg} = -4 \pi G M_{g}\rho_{s}  \int^{R_{s}}_{0}\frac{R^{4}}{R^{3}_{g}}dR $$ 
  one could find $ W^{N}_{sg}$ as
  $$ W^{N}_{sg} =  -\frac{3GM_{s}M_{g}R^{2}_{s}}{5R^{3}_{s}}.  $$

Thus, for the particular mass distribution considered here, the Newtonian share of virial energy due to interaction of stars and gas could be found  as
\begin{eqnarray} \label{WN}
  W^{N}_{gs} +W^{N}_{sg} =   -\frac{3GM_{s}M_{g}}{2R_{g}} \left( 1- \frac{1}{5}\frac{R^{2}_{s}}{R^{2}_{g}}  \right).
\end{eqnarray}

Now, we consider the MOD share of the virial energy of gas due to stars
$$ W^{MOD}_{gs} \equiv \int \rho_{g}d^{3}x~ \vec{x}.\vec{\nabla}\Phi^{MOD}_{s} $$
which could be written as
$$ W^{MOD}_{gs} = \int^{R_{s}}_{0} \rho_{g}d^{3}x~ \vec{x}.\vec{\nabla}\Phi^{MOD}_{s, in} + \int^{R_{g}}_{R_{s}} \rho_{g}d^{3}x~ \vec{x}.\vec{\nabla}\Phi^{MOD}_{s, out}. $$
According to  \cite{Shenavar:2016xcp} (Appendix), the acceleration due to MOD inside   the stellar distribution could be written as  $\vec{\nabla}\Phi^{MOD}_{s, in} = 2c_{1}a_{0}M_{s}\frac{R}{R_{s}}\left[1-\frac{1}{5}(\frac{R}{R_{s}})^{2}    \right]/M$ while for outside the stellar component we have $\vec{\nabla}\Phi^{MOD}_{s, out} = 2c_{1}a_{0}M_{s}\left[1-\frac{1}{5}(\frac{R_{s}}{R})^{2}    \right]/M$. Thus, the previous integral could be written as
 \begin{eqnarray} 
  W^{MOD}_{gs} = -2c_{1}a_{0}\frac{4 \pi M_{s}\rho_{g}}{M}  \int^{R_{s}}_{0}\frac{R^{4}}{R_{s}}\left[1-\frac{1}{5}(\frac{R}{R_{s}})^{2}    \right]  dR    
  - 2c_{1}a_{0}\frac{4 \pi M_{s}\rho_{g}}{M} \int^{R_{g}}_{R_{s}} R^{3}\left[1-\frac{1}{5}(\frac{R_{s}}{R})^{2}    \right] dR
\end{eqnarray} 
which could be evaluated as follows:
$$   W^{MOD}_{gs} = -2c_{1}a_{0}\frac{M_{s}M_{g}}{M}3R_{g} \left( \frac{1}{4} - \frac{1}{10}(\frac{R_{s}}{R_{g}})^{2} + \frac{3}{140}(\frac{R_{s}}{R_{g}})^{4}  \right). $$

Finally, the MOD virial energy of stars due to interaction with gas 
 $$ W^{MOD}_{sg} \equiv \int \rho_{s}d^{3}x~ \vec{x}.\vec{\nabla}\Phi^{MOD}_{g} $$
could be written as
 $$ W^{MOD}_{sg} = \int^{R_{s}}_{0} \rho_{s}d^{3}x~ \vec{x}.\vec{\nabla}\Phi^{MOD}_{g, in}$$  
 which contains only one integral for the same reason as the Newtonian case.  Again, by using the results of the appendix of  \cite{Shenavar:2016xcp} one could find
 $$  W^{MOD}_{sg} = -2c_{1}a_{0} \frac{4 \pi  M_{g}\rho_{s}}{M}  \int^{R_{s}}_{0}\frac{R^{4}}{R_{g}} \left[1-\frac{1}{5}(\frac{R}{R_{g}})^{2}    \right] dR $$
 and solve the integral to derive $  W^{MOD}_{sg}$ as follows:
 $$  W^{MOD}_{sg} = -2c_{1}a_{0} \frac{ M_{g}M_{s}}{M}3R_{g} \left( \frac{1}{5}(\frac{R_{s}}{R_{g}})^{2} - \frac{1}{35}(\frac{R_{s}}{R_{g}})^{4}  \right).  $$ 

Therefore, the MOD share of virial energy due to interaction of stars and gas could be written  as
\begin{eqnarray} \label{WMOD}
 W^{MOD}_{sg} + W^{MOD}_{gs} = -2c_{1}a_{0}\frac{M_{s}M_{g}}{M} \frac{3R_{g}}{4} \left( 1 + \frac{2}{5}(\frac{R_{s}}{R_{g}})^{2} - \frac{1}{35}(\frac{R_{s}}{R_{g}})^{4}  \right).
\end{eqnarray}
In conclusion, for a system in which   both stellar and gas components are  homogeneous and spherical, the total virial energy must be written as 
\begin{eqnarray} 
W_{tot} =& W_{ss} + W_{gg}+ W_{sg} + W_{gs} ~~~~~~~~~~~~~~~~~~~~~~~~~~~~~~~~~~~~~~~~~~~~~~~~~~~~~~~~~~~~~~~~~~~~~~~~~~~~~~~~~~~~~~~~~~~~~~~~~~~~~~~~~~~~~~~~~~~~~~~~~~~~~~~~~~~~~~~~~~~~~~ \\ \nonumber
=& -G \left[ \frac{3}{5} (  \frac{M^{2}_{g}}{R_{g}} + \frac{M^{2}_{s}}{R_{s}})  +\frac{3M_{s}M_{g}}{2R_{g}} \left( 1- \frac{1}{5}\frac{R^{2}_{s}}{R^{2}_{g}}  \right)  \right]  - \frac{2 c_{1}a_{0}  }{M}  \left[ \frac{18}{35} ( M^{2}_{g} R_{g} + M^{2}_{s} R_{s} ) +  \frac{3}{4}  M_{s}M_{g}R_{g} \left( 1 + \frac{2}{5}(\frac{R_{s}}{R_{g}})^{2} - \frac{1}{35}(\frac{R_{s}}{R_{g}})^{4}  \right)   \right] 
\end{eqnarray}
where we have used $\alpha$ and $\beta$ from Table \ref{table:paras}.

\begin{figure} 
\begin{center}
\includegraphics[height=10cm,width=17cm]{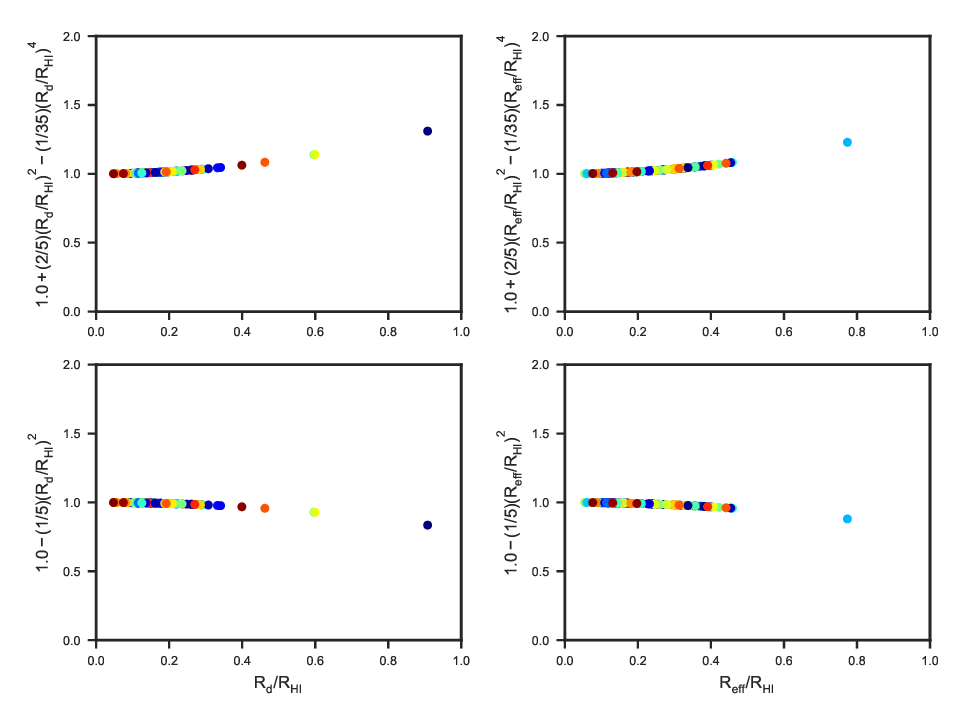} 
  \vspace{-18pt}
  \caption{The terms $1- \frac{1}{5}\frac{R^{2}_{s}}{R^{2}_{g}} $ and $1 + \frac{2}{5}(\frac{R_{s}}{R_{g}})^{2} - \frac{1}{35}(\frac{R_{s}}{R_{g}})^{4}$ are almost constant for most of the data points in our sub-sample of SPARC. Left: It is assumed that $R_{s}=R_{d}$ and $R_{g}=R_{HI}$. Only four data points show a noticeable deviation from one due to their relatively large $R_{d}/R_{HI}$ ratios. Right:   $R_{s}=R_{eff}$. One data point shows about 25 $\%$ deviation from one; the rest show a negligible deviation. }
  \label{fig:omit}
\end{center}
\end{figure}

Now, we will show that the terms in the parentheses  on the RHS of both Eqs. \ref{WN} and \ref{WMOD} are very close to one for most objects in the SPARC sample; thus, they could be approximated as one. These are the sub-leading terms related to the particular mass distributions considered. To see this, we have plotted $1- \frac{1}{5}\frac{R^{2}_{s}}{R^{2}_{g}} $ and $1 + \frac{2}{5}(\frac{R_{s}}{R_{g}})^{2} - \frac{1}{35}(\frac{R_{s}}{R_{g}})^{4}$ in Fig \ref{fig:omit}. If we assume $R_{s}=R_{d}$ and $R_{g}=R_{HI}$, then it could be seen clearly from these two plots that the variation from one is quite negligible for most objects. In fact,  only four objects show a clear deviation from one. For large ratios of $R_{s}/R_{g}$, the variation from unity could be estimated as $35 \%$ at most; however, for smaller ratios of $R_{s}/R_{g}$ the terms are quite close to one. The graph on the right in Fig \ref{fig:omit} shows that  a similar behavior happens when $R_{s}=R_{eff}$.

Therefore, neglecting the higher order terms, i.e. $( R_{s}/ R_{g} )^{n}$ with $n \geq ~2$, the total virial energy of a general system with gaseous and stellar components could be approximated as follows  
\begin{eqnarray}  \nonumber
-W_{tot} \simeq  G \left[ \alpha (  \frac{M^{2}_{g}}{R_{g}} + \frac{M^{2}_{s}}{R_{s}}) + \gamma_{1} \frac{M_{s}M_{g}}{R_{g}}  \right]  + \frac{2 c_{1}a_{0}  }{M}  \left[ \beta ( M^{2}_{g} R_{g} + M^{2}_{s} R_{s} ) +\gamma_{2} M_{s}M_{g} R_{g}  \right] 
\end{eqnarray}
 in which $\gamma_{1}$ and $\gamma_{2}$ are two constants related to the geometry of the two systems.  The terms proportional to $\gamma_{1}$ and $\gamma_{2}$ estimate  the virial interaction of gas and stars up to $35 \%$.   This general form of $ W_{tot} $ is used to derive the equilibrium radius of a compound system above. See Eq. \ref{Wtot} and the discussion thereafter.

\section{Supplementary Figures}
\label{Suppl}
\begin{figure}[H]
\begin{center}
\includegraphics[width=16cm]{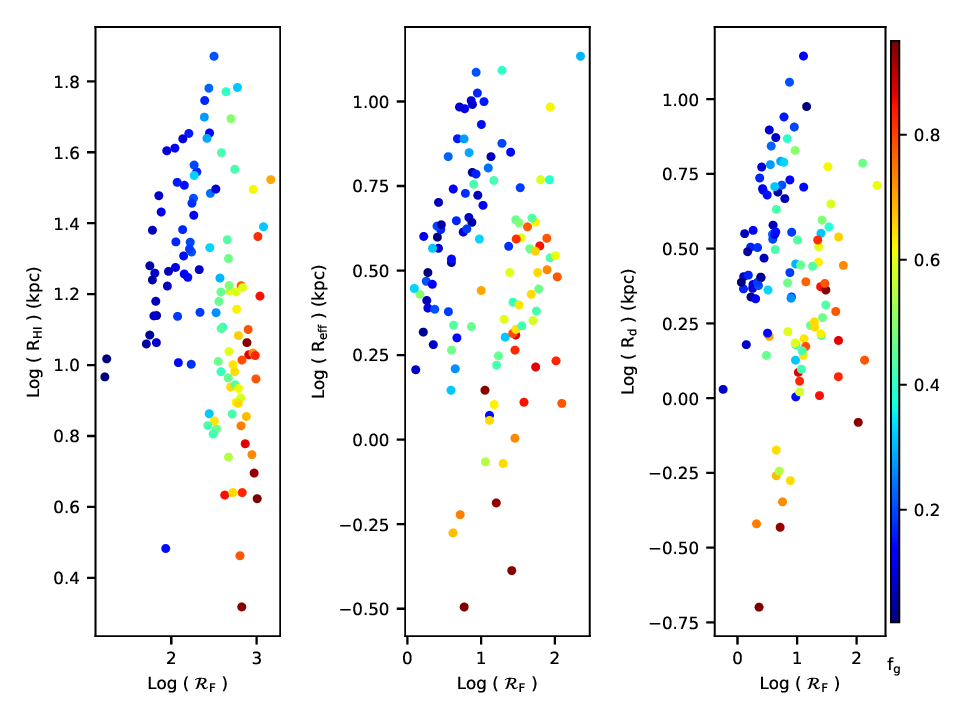}
  \vspace{-25pt}
\caption{ The change in different galactic scales ( $R_{HI},~R_{eff},~R_{d}$ ) as a function  of the Freeman ratio $\mathcal{R}_{F}$. Data from SPARC \citep{2016ApJ...816L..14L,2016AJ....152..157L}. Color online. \label{fig:Freeman}}
\end{center}
\end{figure}

The main results of this work are not dependent to the figures in this appendix; however, we have used these plots in some of our discussions in the manuscript. For example, Fig. \ref{fig:Freeman} displays some scale length, i.e. $R_{HI}$, $R_{eff}$ and $R_{d}$, as a function of the Freeman ratio $\mathcal{R}_{F}$ while we have used $R_{eq}$ as the dynamical scale length of the systems. See also Fig. \ref{fig:FreemanReq}. Other plots have been employed to argue why we prefer  $R_{eq}$ instead of other scale length and $V_{eq}= \sqrt{ 0.5 V^{2}_{f} + \sigma^{2}_{eq} } $ instead of  flat velocity  $V_{f}$.

\begin{figure}
\begin{center}
\vspace{-40pt}
\includegraphics[width=16cm]{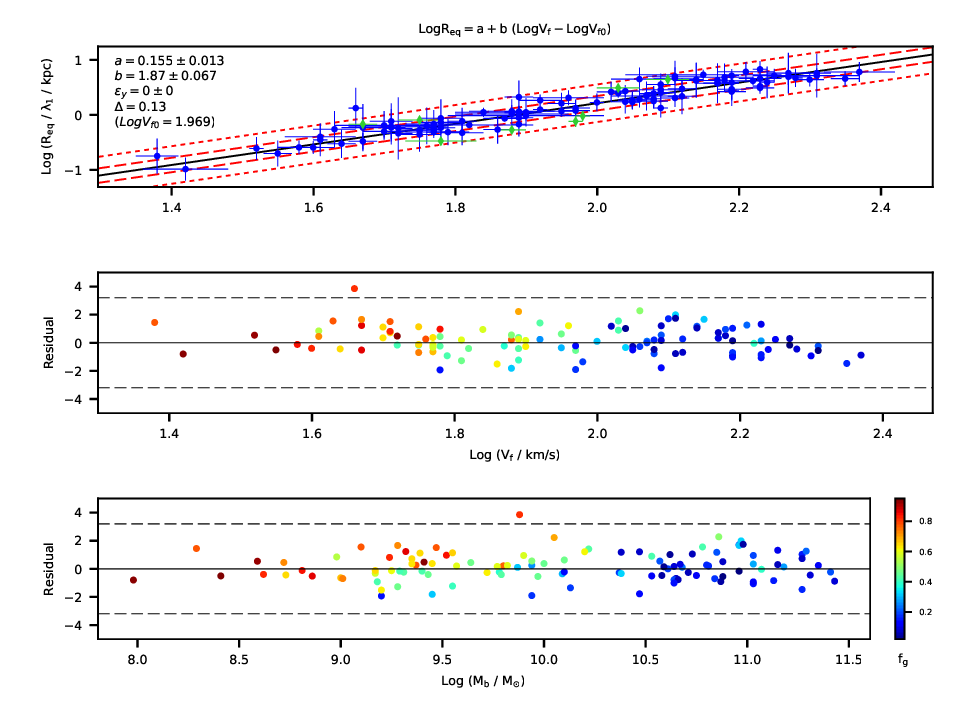}
\vspace{-25pt}
\caption{ \footnotesize{Data analysis of the scaling rule $R_{eq} \propto V^{2}_{eq}$ presuming that $\sigma_{eq}=0$, i.e. in this graph $V_{eq}=V_{f}$. The equilibrium radius $R_{eq}$ is calculated  by assuming $R_{s}=R_{eff}$ and $(\lambda_{2},~ \lambda_{3}, ~ \lambda_{4})/\lambda_{1} = (1.1,~0.039,~0.19)$. It is seen that the slope  is decreased compared to the cases with $\sigma_{eq} \neq 0$. Also, compare with Fig. \ref{fig:V2R}. Data from SPARC \citep{2016ApJ...816L..14L,2016AJ....152..157L}. Color online.} \label{fig:V2Rsig0}}
\end{center}
\end{figure}
\vspace{-50pt}

\begin{figure}
\begin{center}
\vspace{-10pt}
\includegraphics[width=16cm]{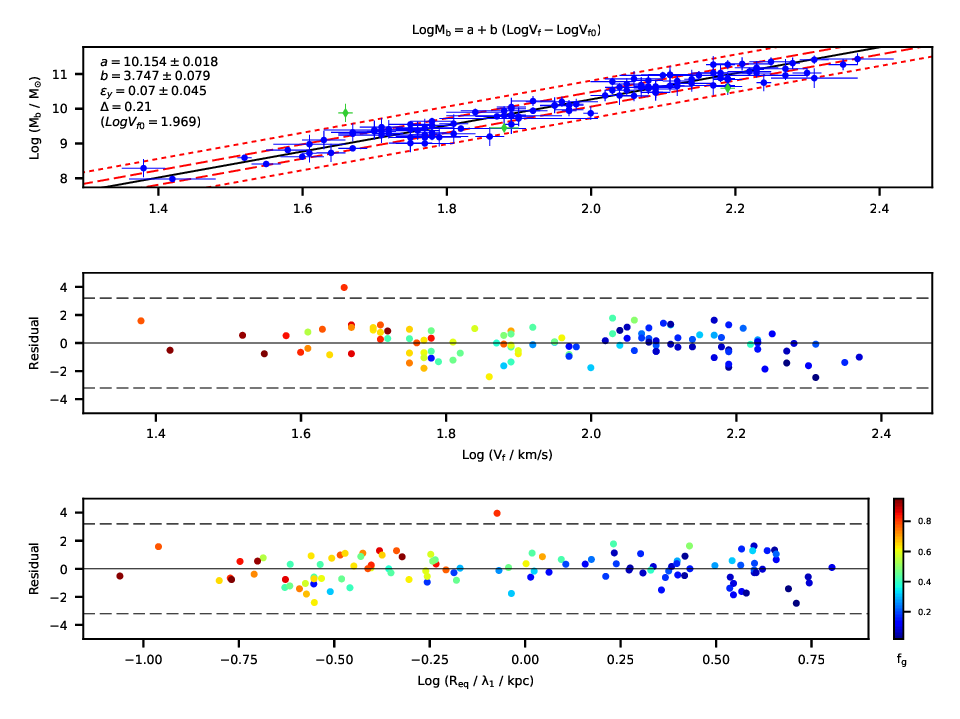} 
  \vspace{-25pt}
\caption{\footnotesize{A decreased slope is observed for  BTFR $M \propto V^{4}_{eq}$ presuming that $\sigma_{eq}=0$, i.e. in this graph $V_{eq}=V_{f}$. The equilibrium radius $R_{eq}$ in lower residual plot  is calculated  by assuming $R_{s}=R_{eff}$ and $(\lambda_{2},~ \lambda_{3}, ~ \lambda_{4})/\lambda_{1} = (1.1,~0.039,~0.19)$. Also, compare with Fig. \ref{fig:TF}.  Data from SPARC \citep{2016ApJ...816L..14L,2016AJ....152..157L}.  Color online.} \label{fig:TFsig0}}
\end{center}
\end{figure}

\begin{figure}
\begin{center}
\vspace{-40pt}
\includegraphics[width=16cm]{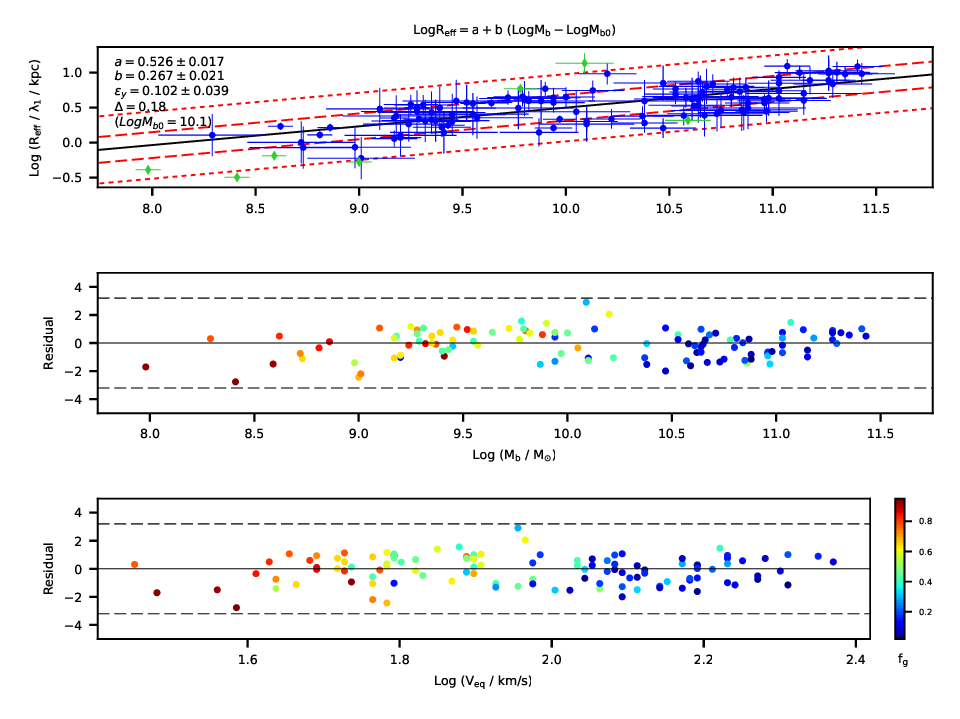}
  \vspace{-25pt}
\caption{ Data analysis of the scaling rule $R_{eff} - M $, i.e. when the scale length of the system is $R_{eff}$ instead of $R_{eq}$, and $\sigma_{eq}=15 ~ km/s$.  Compare with Fig. \ref{fig:R2M}. Data from SPARC \citep{2016ApJ...816L..14L,2016AJ....152..157L}. Color online.  \label{fig:R2MReff}}
\end{center}
\end{figure}

\begin{figure}
\begin{center}
\vspace{-10pt}
\includegraphics[width=16cm]{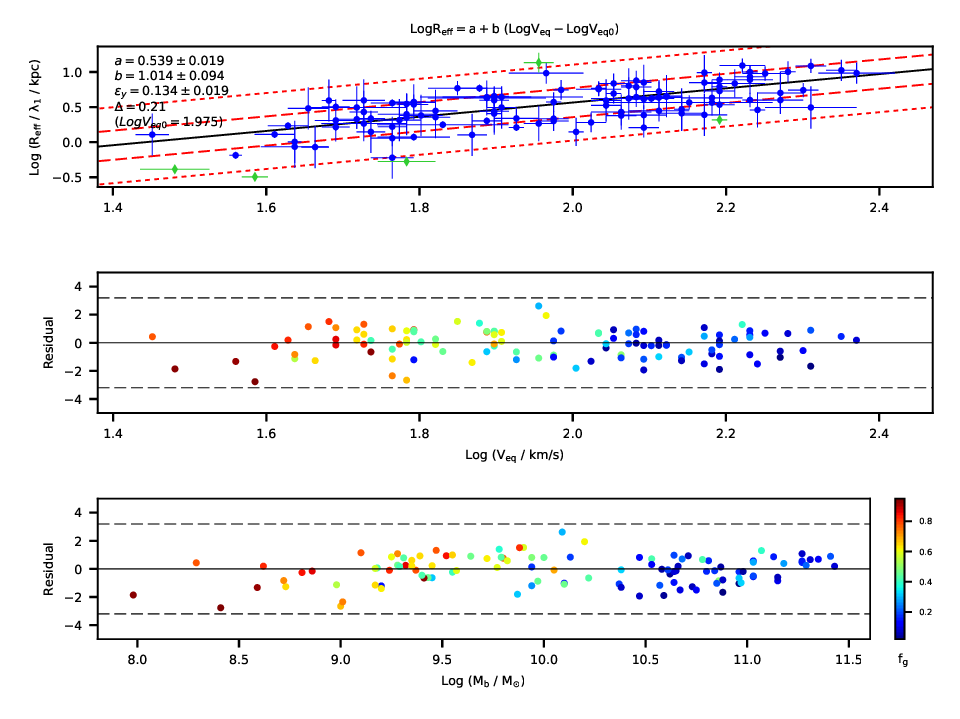}
  \vspace{-25pt}
\caption{ Data analysis of the scaling rule $R_{eff} - V_{eq} $, i.e. when the scale length of the system is $R_{eff}$ instead of $R_{eq}$, and $\sigma_{eq}=15 ~ km/s$.  Compare with Fig. \ref{fig:V2R}. Data from SPARC \citep{2016ApJ...816L..14L,2016AJ....152..157L}.  Color online. \label{fig:V2RReff}}
\end{center}
\end{figure}

\begin{figure}
\begin{center}
\vspace{-40pt}
\includegraphics[width=16cm]{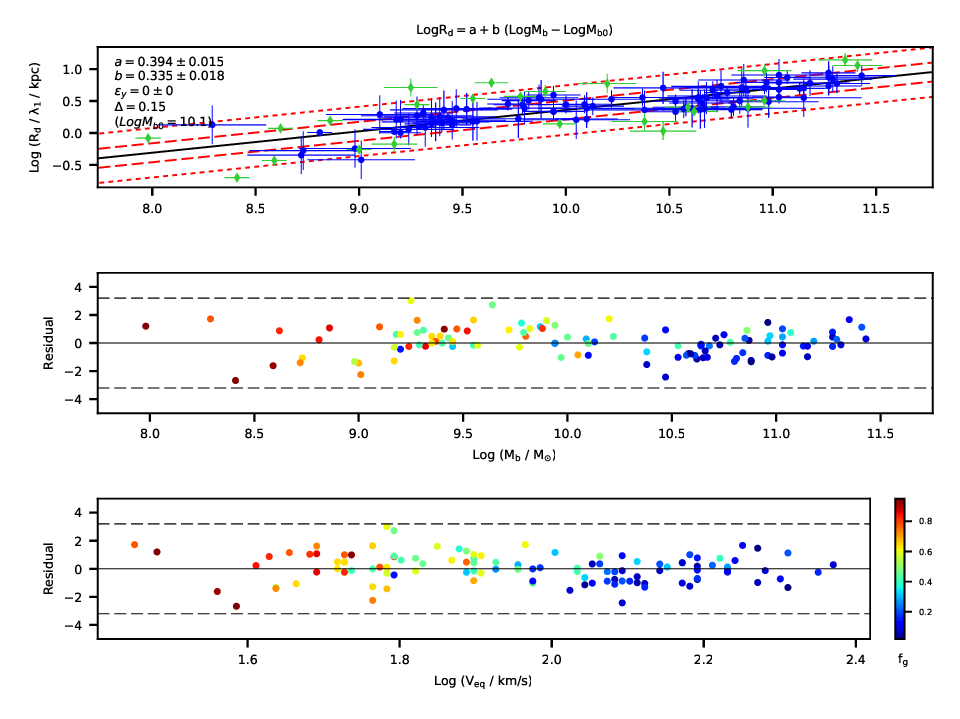}
  \vspace{-25pt}
\caption{ Data analysis of the scaling rule $R_{d} - M $, i.e. when the scale length of the system is $R_{d}$ instead of $R_{eq}$, and $\sigma_{eq}=15 ~ km/s$.  Compare with Fig. \ref{fig:R2M}. Data from SPARC \citep{2016ApJ...816L..14L,2016AJ....152..157L}. Color online. \label{fig:R2MRd}}
\end{center}
\end{figure}

\begin{figure}
\begin{center}
\vspace{-10pt}
\includegraphics[width=16cm]{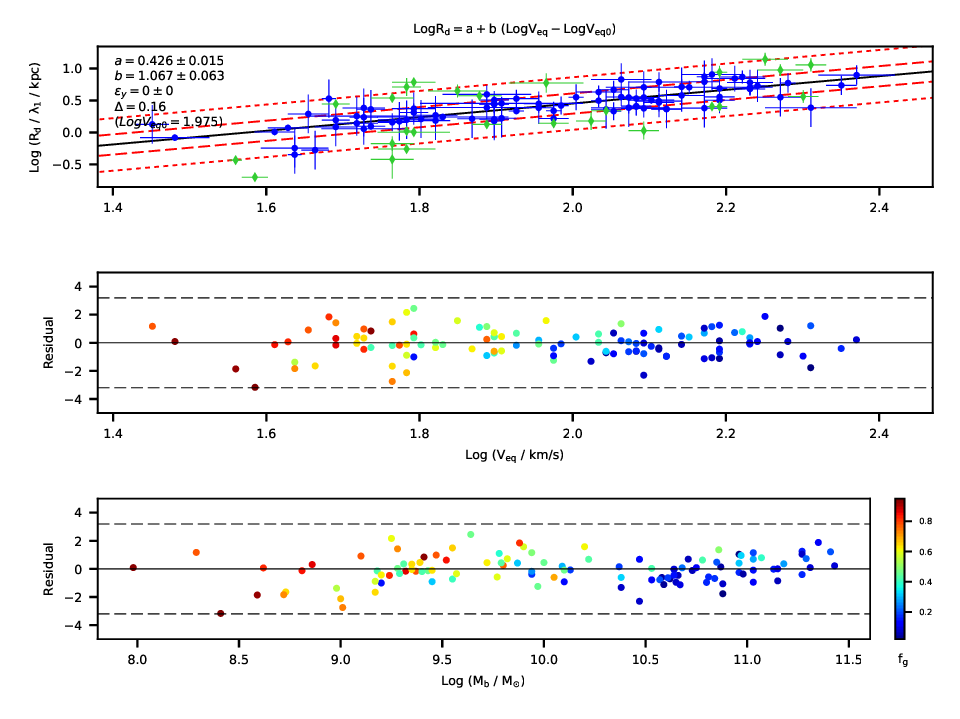}
  \vspace{-25pt}
\caption{ Data analysis of the scaling rule $R_{d} - V_{eq} $, i.e. when the scale length of the system is $R_{d}$ instead of $R_{eq}$, and $\sigma_{eq}=15 ~ km/s$.  Compare with Fig. \ref{fig:V2R}. Data from SPARC \citep{2016ApJ...816L..14L,2016AJ....152..157L}. Color online. \label{fig:V2RRd}}
\end{center}
\end{figure}

\begin{figure}
\begin{center}
\vspace{-40pt}
\includegraphics[width=16cm]{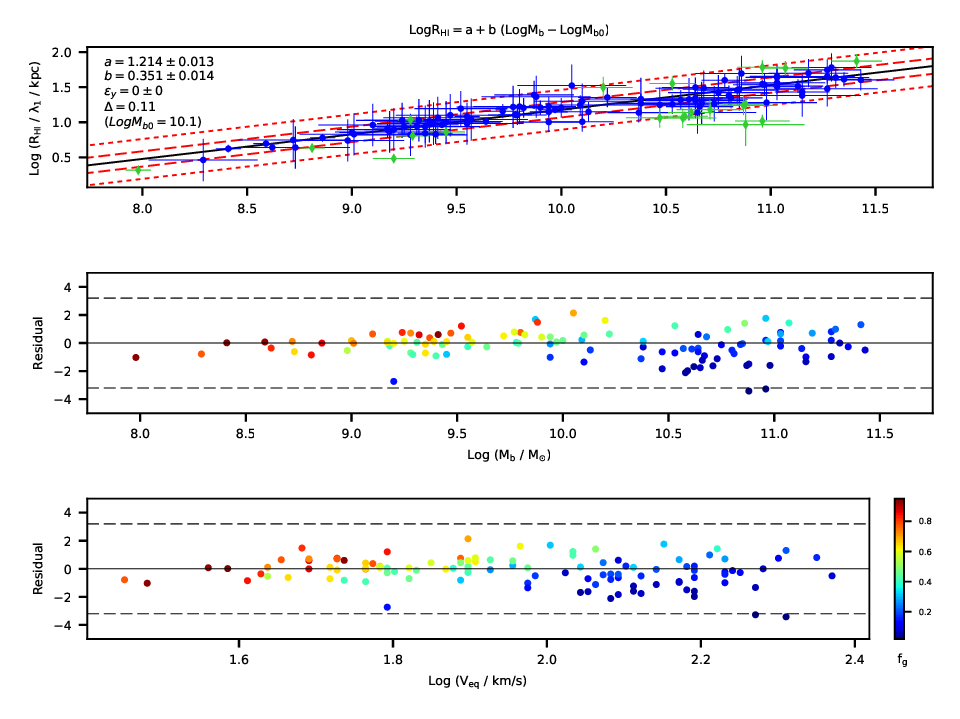}
  \vspace{-25pt}
\caption{ Data analysis of the scaling rule $R_{HI} - M $, i.e. when the scale length of the system is $R_{HI}$ instead of $R_{eq}$, and $\sigma_{eq}=15 ~ km/s$.  Compare with Fig. \ref{fig:R2M}. Data from SPARC \citep{2016ApJ...816L..14L,2016AJ....152..157L}. Color online. \label{fig:R2MRHI}}
\end{center}
\end{figure}

\begin{figure}
\begin{center}
\vspace{-10pt}
\includegraphics[width=16cm]{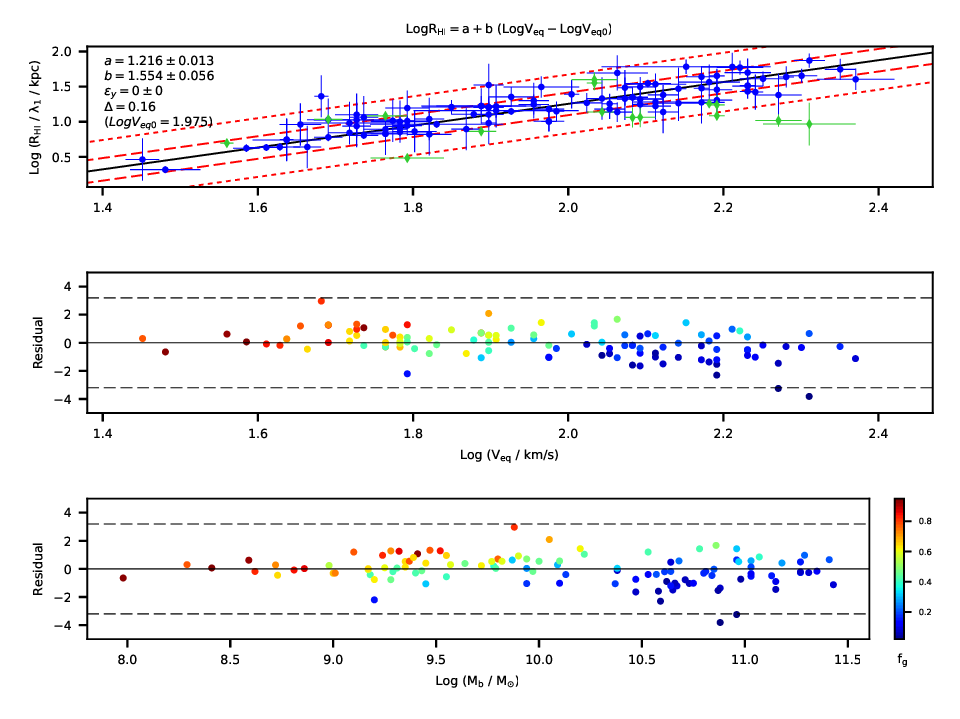}
  \vspace{-25pt}
\caption{ Data analysis of the scaling rule $R_{HI} - V_{eq} $, i.e. when the scale length of the system is $R_{HI}$ instead of $R_{eq}$, and $\sigma_{eq}=15 ~ km/s$.  Compare with Fig. \ref{fig:V2R}. Data from SPARC \citep{2016ApJ...816L..14L,2016AJ....152..157L}. Color online. \label{fig:V2RRHI}}
\end{center}
\end{figure}

\vspace{40pt}
\bibliographystyle{plainnat}
\bibliography{scaling}

\end{document}